\documentclass[12pt,draftclsnofoot,onecolumn]{IEEEtran}
\usepackage{caption2}
\usepackage{graphicx}
\usepackage{subfigure}
\usepackage{booktabs}
\usepackage{multirow}
\usepackage{verbatim}
\usepackage{subeqnarray}
\usepackage{bm}
\usepackage{rotating}
\usepackage{algorithmicx,algorithm}
\usepackage{algpseudocode,algorithm}
\usepackage{cite}%
\usepackage[
            linkcolor=red,
            anchorcolor=blue,
            citecolor=green]{hyperref}
\usepackage{amsmath}
\usepackage{amssymb}
\usepackage{booktabs}
\usepackage{color}
\usepackage[amsmath,thmmarks]{ntheorem}
\usepackage{array}
\renewcommand\arraystretch{1}
\theoremseparator{:}

\newtheorem{theorem}{Theorem}

\newtheorem{corollary}{Corollary}
\newtheorem{example}{Example}

\newcounter{RomanNumber}
\newcommand{\MyRoman}[1]{\setcounter{RomanNumber}{#1}\Roman{RomanNumber}}

\ifCLASSINFOpdf

\else

\fi

\begin{document}
%
\title{Efficient File Delivery for Coded Prefetching in Shared Cache Networks with Multiple Requests Per User}


\author{
Haisheng Xu,~\IEEEmembership{Member,~IEEE},
 ~Chen Gong,~\IEEEmembership{Member,~IEEE},
 ~and Xiaodong Wang,~\IEEEmembership{Fellow,~IEEE}
\thanks{H. Xu  and X. Wang are with the Department of Electrical Engineering, Columbia University, New York, NY 10027, USA. (e-mail: hx2219@columbia.edu, wangx@ee.columbia.edu).
\par C. Gong is with the Key Laboratory of Wireless Optical
Communications, Chinese Academy of Sciences, School of Information
Science and Technology, University of Science and Technology
of China, Hefei 230027, China (e-mail: cgong821@ustc.edu.cn).}}

\markboth{ }%
{Shell \MakeLowercase{\textit{et al.}}: Bare Demo of IEEEtran.cls for Journals}
\IEEEtitleabstractindextext{%
\begin{abstract}
We consider a centralized caching network, where a server serves several groups of users, each having a common shared homogeneous fixed-size cache and requesting arbitrary multiple files. An existing coded prefetching scheme is employed where each file is broken into multiple fragments and each cache stores multiple coded packets each formed by XORing fragments from different files. For such a system, we propose an efficient file delivery scheme with explicit constructions by the server to meet the arbitrary multi-requests of all user-groups. Specifically, the stored coded packets of each cache are classified into four types based on the composition of the file fragments encoded. A delivery strategy is developed, which separately delivers part of each packet type first and then combinatorially delivers the remaining different packet types in the last stage.  The rate as well as the worst rate of the proposed delivery scheme are analyzed. We show that our caching model and delivery scheme can incorporate some existing coded caching schemes as special cases. Moreover, for the special case of uniform requests and uncoded prefetching, we make a comparison with existing results, and show that our approach can achieve a lower delivery rate. We also provide numerical results on the delivery rate for the proposed scheme.
\end{abstract}
}

\maketitle

\IEEEdisplaynontitleabstractindextext

\IEEEpeerreviewmaketitle

\section{Introduction}
Coded caching \cite{Maddah-Ali2014,Li2017,Maddah-Ali2015,Niesen2017,Amiri2017,Ji2016a} is a technique that can reduce the communication load or latency in data distribution via proactively storing part of data to users' caches during off-peak traffic periods and thus transmitting less amount of data to users during peak traffic periods. The transmissions typically consist of two phases: \emph{placement (or prefetching) phase} and \emph{delivery (or transmission) phase}. In the placement phase, uncoded \cite{Maddah-Ali2014,Wan2016,Yu2018} or coded \cite{Bioglio2015,Chen2016,Tian2018,Amiri2017,Zhang2018,Wei2017,Gomez-Vilardebo2016} partial contents are placed into each user's cache without knowing the future requests. According to whether the users' caches are coordinated or not in the placement phase, coded caching can be divided into \emph{centralized} \cite{Maddah-Ali2014} and \emph{decentralized} \cite{Maddah-Ali2015} settings, where the latter adopts independent and identical random prefetching strategies across caches. Since all users' cached contents are coordinately placed in a deterministic way by the server, centralized coded caching shows a lower delivery rate than that of decentralized coded caching. Existing placement strategies in the centralized settings are categorized as uncoded and coded prefetching, respectively. Under coded prefetching, the storage resources of the caching network can be more efficiently utilized. Although such higher storage efficiency may be obtained at the cost of delivery rate increment, coded prefetching can achieve an improved trade-off between the (worst) delivery rate and the cache memory size than uncoded prefetching when the cache size is relatively small\cite{Tian2018,Gomez-Vilardebo2018}.
\par A number of the centralized coded caching algorithms with coded prefetching have been proposed. In particular, \cite{Chen2016} first proposed a coded prefetching scheme using binary-addition (XOR) codes for the point that each cache size $C$ is the inverse of the number of the users $K$, i.e., $C=\frac{1}{K}$. Then \cite{Tian2018} proposed a coded prefetching scheme using a combination of rank metric codes and maximum distance separable (MDS) codes for $K$ cache-size points at $C=\frac{(K-1)r+(N-1)(r-1)r}{K(K-1)}$, $r=1,2,...,K$, which locate in the regime when each cache size is not greater than the total source-file size, i.e., $0\leq C\leq N$ with $N$ denoting the total file number. Later \cite{Zhang2018} showed that such codes used in \cite{Tian2018} can be simply replaced by the XOR codes. It is shown that the rate-memory pair of \cite{Chen2016} can be viewed as a special case of \cite{Tian2018}, and according to \cite{Zhang2018} the scheme in \cite{Tian2018} can outperform that in \cite{Yu2018} within the small cache-size regime when the total cache size of the network is less than the total source-file size, i.e., $0\leq C<\frac{N}{K}$, where \cite{Yu2018} proposed an uncoded prefetching scheme for $K$ cache-size points at $C=\frac{tN}{K}$, $t=1,2,...,K$ over $0\leq C\leq N$ based on \cite{Maddah-Ali2014} and is shown to be optimal in the regime when $\frac{N}{K}\leq C\leq N$. Since coded prefetching can achieve better performance in small cache-size regime, \cite{Amiri2017} proposed a coded prefetching scheme for a cache-size point at $C=\frac{N-1}{K}$ and \cite{Gomez-Vilardebo2016} proposed a coded prefetching scheme for $N$ more cache-size points at $C=\frac{N}{K\alpha},~\alpha=1,2,...,N$ over $0\leq C\leq \frac{N}{K}$. It is shown that \cite{Gomez-Vilardebo2016} can include coded prefetching \cite{Chen2016,Tian2018} at $C=\frac{1}{K}$ and uncoded prefetching \cite{Yu2018} at $C=\frac{N}{K}$ as special cases, and can further improve coded prefetching performance over such small cache-size regime \cite{Tian2018,Gomez-Vilardebo2018}. However, all the aforementioned coded prefetching schemes are only applicable to the one-user-per-cache network, where each user can only make a single request. To the best of our knowledge, there are few works considering coded prefetching for multiple requests. Note that \cite{Ji2014,Sengupta2017} investigated single-layer coded caching with multiple requests and \cite{Zhang2016,Karamchandani2016} investigated hierarchical coded caching with multiple requests, all of which address uniform requests for uncoded prefetching. Moreover, \cite{Zhang2016} focused on centralized hierarchical uncoded prefetching that can degenerate into \cite{Ji2014,Sengupta2017} with the same delivery rate, while \cite{Karamchandani2016} focused on decentralized hierarchical uncoded prefetching.
 \par In this paper, we focus on centralized coded prefetching with small sum-size caches for arbitrary multiple
requests per user in the regime when the total cache size is not greater than the total source-file size of the server. Consider a caching network consisting of one server and several groups of users, where each group has a common shared fixed-size cache with the size homogeneously allocated over different groups and each user in a user-group can make arbitrary multiple requests for the files stored in the server. The caching model can degenerate into existing coded prefetching \cite{Chen2016,Tian2018,Gomez-Vilardebo2016,Zhang2018,Amiri2017,Gomez-Vilardebo2018} when each user-group consists of only one user and makes a single request, but more generally we can consider arbitrary multiple requests for each user. Our model is motivated by FemtoCaching networks \cite{Golrezaei2013,Shanmugam2013} and cache-enabled small-cell networks \cite{Bioglio2015,Xu2017}, where a number of homogeneous cache-enabled small-cell base stations receive data from a controlling macro base station via a cellular downlink, and each small-cell base station serves a group of users through its own local high-rate downlink. As a prefetching scheme does not depend on the specific requests of users and our paper focuses on the same cache-size regime with that of \cite{Gomez-Vilardebo2016}, we cache coded contents based on the prefetching scheme given in \cite{Gomez-Vilardebo2016}, where each file is broken into multiple fragments and each cache stores multiple coded packets each formed by XORing fragments from different files. Then an efficient file delivery scheme with explicit constructions by the server to meet the multi-requests of all user-groups is proposed. The delivery scheme is more complete and general than that of \cite{Gomez-Vilardebo2016} such that each user in a group can request arbitrary multiple files. Specifically, the stored coded packets of each cache are classified into four types based on the composition of the file fragments encoded. A delivery strategy is developed by separately delivering part of each packet type first and then combinatorially delivering the remaining different packet types in the last stage. After that, the rate as well as the worst rate of the proposed delivery scheme are analyzed since under our delivery scheme we can not only calculate the worst delivery rate as usually provided by \cite{Ji2014,Sengupta2017,Zhang2016,Karamchandani2016} but also calculate the actual delivery rate for each group of specific multiple requests. We show that our caching model and delivery scheme can incorporate \cite{Gomez-Vilardebo2016} as a very special case, which includes some cases of the coded\cite{Chen2016,Tian2018} and uncoded \cite{Yu2018} prefetching as mentioned above. Moreover, for the special case of uniform requests and uncoded prefetching, we make a comparison with existing schemes \cite{Ji2014,Sengupta2017} at the same cache size point, and show that our approach can achieve a lower delivery rate. Finally, the performance of the proposed delivery scheme is numerically evaluated.
 \par The remainder of this paper is organized as follows. In Section \ref{sec:SystemModel}, the cache network model is given and the coded prefetching scheme is described. In Sections \ref{sec:CodedcachingScheme}, the delivery scheme is proposed. In Section \ref{sec:Analyses}, we provide the analyses on the delivery rate and worst delivery rate for the proposed delivery scheme. Numerical results are provided in Section \ref{sec:NumResults}. Finally, Section \ref{sec:Conclusion} concludes the paper.
 \begin{figure*}[ht]
  \centering
  {\begin{minipage}{1\textwidth}
  \centering
  \includegraphics[scale=0.64]{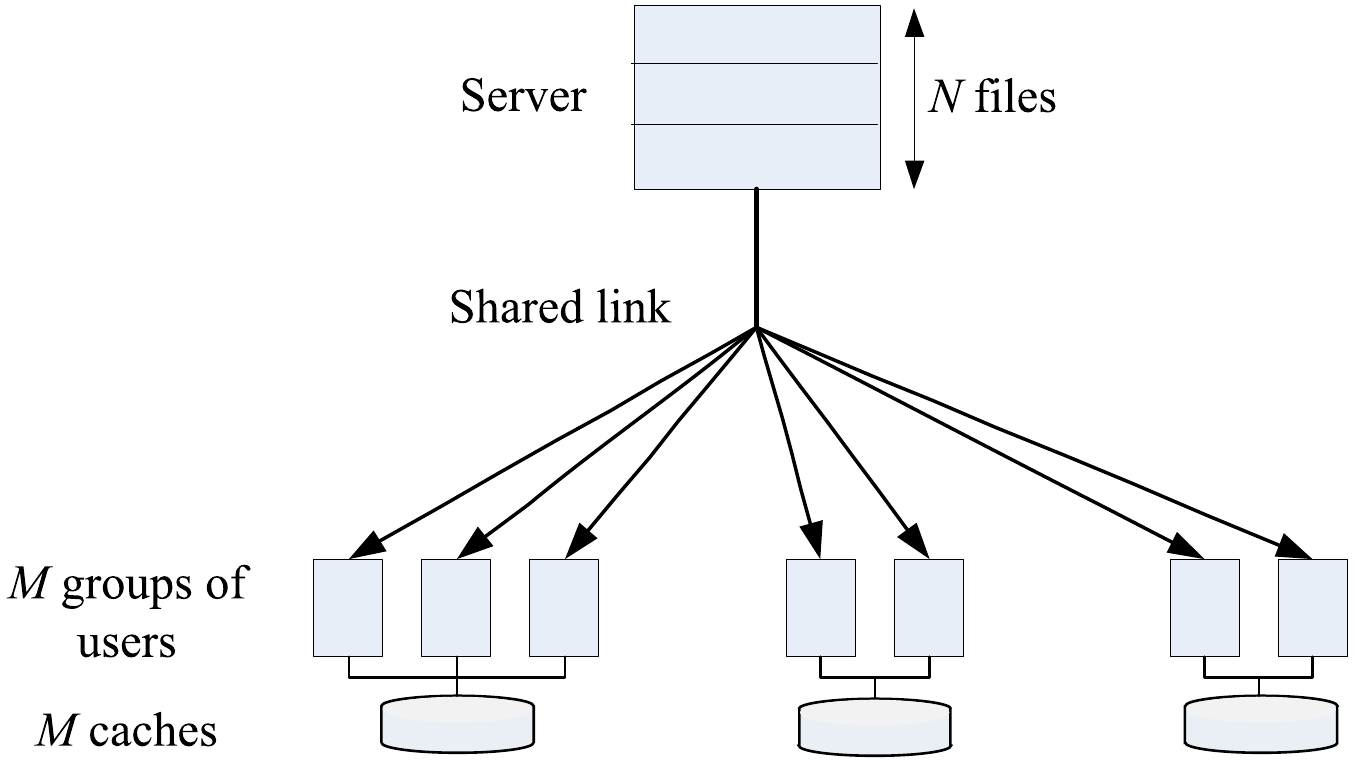}\label{subfig:DC_Architecture1}
  \end{minipage}  }
  \caption{A shared-cache network.} \label{fig:Fig1_CCA}
\end{figure*}
\section{Background}\label{sec:SystemModel}
\subsection{System Description}
Consider a centralized cache network as shown in Fig. \ref{fig:Fig1_CCA}, which consists of one server and several groups of users each sharing a common equal-size cache. This network is characterized by parameters $(N,M,D,\alpha)$ as follows:
\begin{itemize}
  \item The server has a database of $N$ unit-size files $S_{1},S_{2},\ldots,S_{N}$. Denote the file index set by $\mathcal{N}\triangleq\{1,2,...,N\}$.
  \item There are $M$ groups of users connected to the server via a shared error-free link \cite{Maddah-Ali2014}. Denote the user-group index set by $\mathcal{M}\triangleq\{1,2,...,M\}$. Each user-group requests files from the server with the assistance of its shared cache.
  \item Let $D$ denote the sum of all user-groups' distinct request numbers, which satisfies $M\leq D\leq NM$. Assume user-group $m\in\mathcal{M}$ has $D_{m}\geq 1$ distinct requests denoted as $\mathcal{D}_{m}$, then we have $\sum\limits_{m=1}^{M} D_{m}=D$ and $\bigcup\limits_{m=1}^{M}\mathcal{D}_{m}\subseteq \mathcal{N}$.
  \item $\alpha=\frac{N}{M C}$ is a memory-size parameter, where $C$ is the size of each cache, $1\leq \alpha\leq N$ and thus the cache size $C$ satisfies $\frac{1}{M}\leq C\leq \frac{N}{M}$. Note that when $\alpha=1$ we have uncoded prefetching and otherwise we have coded prefetching.
\end{itemize}
\par Note also that when each user-group consists of only one user and requests the same number of distinct files, the above $(N,M,D,\alpha)$ coded caching network becomes the traditional one-user-per-cache network with $D=M$ \cite{Chen2016,Tian2018,Amiri2017,Gomez-Vilardebo2016} or with $D_{1}=D_{2}=\cdots=D_{M}\geq1$ \cite{Ji2014,Sengupta2017}.
 \subsection{Coded Prefetching Scheme}
 We adopt the coded prefetching scheme in \cite{Gomez-Vilardebo2016} consisting of a cached content assignment step and an assigned content coding step, as follows:
 \begin{itemize}
   \item Step 1: Split each file $S_{n}$, $n\in \mathcal{N}$, into $M\textsf{\textit{C}}_{N-1}^{\alpha-1}$ non-overlapping fragments of equal size and then assign $\textit{\textsf{C}}_{N-1}^{\alpha-1}$ fragments to cache $m$ for any $m\in\mathcal{M}$. Then cache $m$ is assigned with $N\textsf{\textit{C}}_{N-1}^{\alpha-1}$ distinct fragments.
   \item Step 2: Perform XOR among each combination of $\alpha$ fragments each from a different file. For $N$ files, the number of such combinations is $\textsf{\textit{C}}_{N}^{\alpha}$ with the combination set defined as
       \begin{equation}\label{equ:CodedPrefetching:Definition1}
         \mathcal{A}\triangleq \left\{(n_{1},...,n_{\alpha}):~ n_{1}< n_{2}<\cdots< n_{\alpha},~n_{i}\in\mathcal{N}\right\};
       \end{equation}
        and each file occurs in $\textsf{\textit{C}}_{N-1}^{\alpha-1}$ combinations with the combination set containing file $S_{n}$ defined as
        \begin{equation}\label{equ:CodedPrefetching:Definition2}
         \mathcal{A}_{n}\triangleq \left\{(n_{1},...,n_{\alpha})\in \mathcal{A}~\mathrm{such}~\mathrm{that}~n_{j}=n~\mathrm{for}~\mathrm{some}~1\leq j\leq\alpha \right\}.
       \end{equation}
       Then we can index the $\textsf{\textit{C}}_{N-1}^{\alpha-1}$ fragments of $S_{n}$ assigned to cache $m$ as $S_{n,(n_{1},...,n_{\alpha})}^{(m)}$, $(n_{1},...,n_{\alpha})\in\mathcal{A}_{n}$. Thus the $\textit{\textsf{C}}_{N}^{\alpha}$ cached packets in cache $m$ are
\begin{equation}\label{equ:CachedPacketsSet1}
 P_{(n_{1},...,n_{\alpha})}^{(m)}=\bigoplus\limits_{i=1}^{\alpha} S_{n_{i},(n_{1},...,n_{\alpha})}^{(m)},~(n_{1},...,n_{\alpha})\in\mathcal{A}.
\end{equation}
 \end{itemize}
 \par As the size of each packet is $\frac{1}{M\textsf{\textit{C}}_{N-1}^{\alpha-1}}$, the cache size of each user-group is given by $C=\frac{\textsf{\textit{C}}_{N}^{\alpha} }{M\textsf{\textit{C}}_{N-1}^{\alpha-1}}=\frac{N}{\alpha M}$. We illustrate the coded prefetching through the following example.
\begin{example}
Consider $(N,M,\alpha)=(3,3,2)$. The cache size is $C=\frac{N}{M\alpha}=\frac{1}{2}$. Splitting each file into $M\textit{\textsf{C}}_{N-1}^{\alpha-1}=6$ non-overlapping fragments and storing every $\textit{\textsf{C}}_{N-1}^{\alpha-1}=2$ fragments in cache $m\in\{1,2,3\}$, then the cached packets are shown in Table \ref{tab:CodedCachingContents_HCMS}.
\end{example}
\begin{table} [ht]
  \centering
\begin{tabular}{|c|c|c|c|c|}
  \hline
  \multicolumn{2}{|c|}{User-group $m$}&1&2&3  \\
  \hline
  \multicolumn{2}{|c|}{$\mathcal{A}$}&$(1,2)$,~$(1,3)$,~$(2,3)$&$(1,2)$,~$(1,3)$,~$(2,3)$&$(1,2)$,~$(1,3)$,~$(2,3)$ \\
  \hline
  \multirow{6}{*}{Prefetching}&\multirow{2}{*}{Assigned fragments}&$S_{1,(1,2)}^{(1)}$,~$S_{2,(1,2)}^{(1)}$&$S_{1,(1,2)}^{(2)}$, $S_{2,(1,2)}^{(2)}$&$S_{1,(1,2)}^{(3)}$, $S_{2,(1,2)}^{(3)}$ \\
  &\multirow{2}{*}{$S_{n,(n_{1},n_{2})\in\mathcal{A}_{n}}^{(m)}$}&$S_{1,(1,3)}^{(1)}$,~$S_{3,(1,3)}^{(1)}$&$S_{1,(1,3)}^{(2)}$,~$S_{3,(1,3)}^{(2)}$&$S_{1,(1,3)}^{(3)}$,~$S_{3,(1,3)}^{(3)}$\\
   &&$S_{2,(2,3)}^{(1)}$,~$S_{3,(2,3)}^{(1)}$&$S_{2,(2,3)}^{(2)}$,~$S_{3,(2,3)}^{(2)}$&$S_{2,(2,3)}^{(3)}$,~$S_{3,(2,3)}^{(3)}$\\
   \cline{2-5}
   &\multirow{2}{*}{All cached packets}&$S_{1,(1,2)}^{(1)}\oplus S_{2,(1,2)}^{(1)}$&$S_{1,(1,2)}^{(2)}\oplus S_{2,(1,2)}^{(2)}$&$S_{1,(1,2)}^{(3)}\oplus S_{2,(1,2)}^{(3)}$\\
   &\multirow{2}{*}{$P_{(n_{1},n_{2})\in\mathcal{A}}^{(m)}$}&$S_{1,(1,3)}^{(1)}\oplus S_{3,(1,3)}^{(1)}$&$S_{1,(1,3)}^{(2)}\oplus S_{3,(1,3)}^{(2)}$&$S_{1,(1,3)}^{(3)}\oplus S_{3,(1,3)}^{(3)}$ \\
   &&$S_{2,(2,3)}^{(1)}\oplus S_{3,(2,3)}^{(1)}$&$S_{2,(2,3)}^{(2)}\oplus S_{3,(2,3)}^{(2)}$&$S_{2,(2,3)}^{(3)}\oplus S_{3,(2,3)}^{(3)}$ \\
  \hline
\end{tabular}\caption{Illustration of coded prefetching.}\label{tab:CodedCachingContents_HCMS}
\end{table}
\section{Proposed Delivery Scheme}\label{sec:CodedcachingScheme}
 Define $N_{\mathrm{R}}$ ($\alpha\leq N_{\mathrm{R}}\leq N$) as the total number of distinct files requested by the users in the whole network and $\mathcal{N}_{\mathrm{R}}$ as the corresponding requested file index set. Thus given the requests $\mathcal{D}_{1}$, $\mathcal{D}_{2}$,..., $\mathcal{D}_{M}$, we have $\mathcal{N}_{\mathrm{R}}=\bigcup\limits_{m=1}^{M}\mathcal{D}_{m}$. Define the \emph{requested fragments} as the fragments from the files in $\mathcal{N}_{\mathrm{R}}$, i.e., $\left\{S_{n,(n_{1},...,n_{\alpha})\in\mathcal{A}_{n}}^{(m\in\mathcal{M})}:~n\in\mathcal{N}_{\mathrm{R}}\right\}$ and the \emph{unrequested fragments} as the fragments from any files in $\mathcal{N}\setminus\mathcal{N}_{\mathrm{R}}$, , i.e., $\left\{S_{n,(n_{1},...,n_{\alpha})\in\mathcal{A}_{n}}^{(m\in\mathcal{M})}:~n\in\mathcal{N}\setminus\mathcal{N}_{\mathrm{R}}\right\}$. Then our goal during the file delivery stage is to deliver the requested fragments $\left\{S_{n,(n_{1},...,n_{\alpha})\in\mathcal{A}_{n}}^{(m\in\mathcal{M})}:~n\in\mathcal{D}_{m}\right\}$ to user-group $m$. We classify the cached packets $\{P_{(n_{1},...,n_{\alpha})\in\mathcal{A}}^{(m)}:~m\in\mathcal{M}\}$ into four types according to the composition of the file fragments encoded, and devise the delivery strategy for the requested fragments in these packets according to the packet type.
\subsection{Type\emph{-\MyRoman{1}} Cached Packets}\label{sec:subsec:DeliveryschemeforType1}
A cached packet $P_{(n_{1},...,n_{\alpha})}^{(m)}$ is Type-\MyRoman{1} if it is encoded from both requested and unrequested fragments, i.e.,
\begin{equation}\label{equ:def:PacketofType1}
  P_{(n_{1},...,n_{\alpha})}^{(m)}=\bigoplus\limits_{i=1}^{\alpha} S_{n_{i},(n_{1},...,n_{\alpha})}^{(m)},~\exists~n_{i_{1}}, n_{i_{2}}\in\{n_{1},...,n_{\alpha}\}~\mathrm{such}~\mathrm{that}~n_{i_{1}}\in\mathcal{N}_{\mathrm{R}},n_{i_{2}}\in\mathcal{N}\setminus\mathcal{N}_{\mathrm{R}}.
\end{equation}
For each Type-\MyRoman{1} packet $P_{(n_{1},...,n_{\alpha})}^{(m)}$, the server directly transmits the requested fragments encoded in it, i.e.,
\begin{equation}\label{equ:TransmittedContents_Type1}
S_{n_{i},(n_{1},...,n_{\alpha})}^{(m)},~n_{i}\in\mathcal{N}_{\mathrm{R}}.
\end{equation}
\par To illustrate this, we use an example based on the cached contents given in Table \ref{tab:CodedCachingContents_HCMS} by assuming $\mathcal{D}_{1}=\{1,2\}$, $\mathcal{D}_{2}=\{2\}$ and $\mathcal{D}_{3}=\{1,2\}$. Then Type-\MyRoman{1} packets in the three caches are $S_{1,(1,3)}^{(m)}\oplus S_{3,(1,3)}^{(m)}$ and $S_{2,(2,3)}^{(m)}\oplus S_{3,(2,3)}^{(m)}$, $m=1,2,3$ and the server just transmits $S_{1,(1,3)}^{(m)}$ and $S_{2,(2,3)}^{(m)}$, $m=1,2,3$. The delivery load is 6 fragments.
\par We now theoretically compute the delivery load according to Type-\MyRoman{1} packets. For any $n_{i}\in \mathcal{N}_{\mathrm{R}}$, all the fragments of $S_{n_{i}}$ assigned to cache $m$ are $\{S_{n_{i},(n_{1},...,n_{\alpha})}^{(m)}:(n_{1},...,n_{\alpha})\in\mathcal{A}_{n_{i}}\}$ with the total number being $\textsf{\textit{C}}_{N-1}^{\alpha-1}$ and the number of the fragments such that $(n_{1},...,n_{\alpha})\in\mathcal{N}_{\mathrm{R}}$ being $\textit{\textsf{C}}_{N_{\mathrm{R}}-1}^{\alpha-1}$. Thus the number of the fragments such that $\{n_{1},...,n_{\alpha}\}\cap(\mathcal{N}\setminus\mathcal{N}_{\mathrm{R}})\neq\emptyset$ is $\textit{\textsf{C}}_{N-1}^{\alpha-1}-\textit{\textsf{C}}_{N_{\mathrm{R}}-1}^{\alpha-1}$, which equals to the number of the assigned fragments of $S_{n_{i}}$ encoded into Type-\MyRoman{1} packets in cache $m$. Since there are $M$ caches and $N_{\mathrm{R}}$ distinct files requested by the users, the number of the transmitted fragments according to Type-\MyRoman{1} packets is given by
     \begin{equation}\label{equ:NumofBroadcastforType1_HMS}
     \begin{split}
       T_{\mathrm{\MyRoman{1}}}&=MN_{\mathrm{R}}\left(\textit{\textsf{C}}_{N-1}^{\alpha-1}-\textit{\textsf{C}}_{N_{\mathrm{R}}-1}^{\alpha-1}\right).
     \end{split}
     \end{equation}
\subsection{Type\emph{-\MyRoman{2}} Cached Packets}
A cached packet $P_{(n_{1},...,n_{\alpha})}^{(m)}$ is Type-\MyRoman{2} if it is encoded by requested fragments only, among which only one is requested by the user-group that caches it. Define the combination set of every $\alpha$ requested fragments each from a different file from $\mathcal{N}_{\mathrm{R}}$ as
\begin{equation}\label{equ:CodedPrefetching:Definition3}
\widetilde{\mathcal{A}}\triangleq \left\{(n_{1},...,n_{\alpha}):~n_{1}< n_{2}<\cdots< n_{\alpha},~n_{i}\in\mathcal{N}_{\mathrm{R}}\right\},
       \end{equation}
where $|\widetilde{\mathcal{A}}|=\textit{\textsf{C}}_{N_{\mathrm{R}}}^{\alpha}$.
Then Type-\MyRoman{2} packets in cache $m$ can be characterized by
 \begin{equation}\label{equ:Packets_Type2}
   P_{(n_{1},...,n_{\alpha})}^{(m)}=\bigoplus\limits_{i=1}^{\alpha} S_{n_{i},(n_{1},...,n_{\alpha})}^{(m)},~ (n_{1},...,n_{\alpha})\in\widetilde{\mathcal{A}}~\mathrm{and}~\left|\{n_{1},...,n_{\alpha}\} \cap \mathcal{D}_{m}\right|=1.
 \end{equation}
Suppose $n_{i}\in\mathcal{D}_{m}$, then we call the fragment $S_{n_{i},(n_{1},...,n_{\alpha})}^{(m)}$ the \emph{local fragment} since it may not need to be transmitted during the delivery.
We first use the following example to illustrate Type-\MyRoman{2} packets and the delivery scheme according to them.
 \begin{table*} [ht]
  \centering
\begin{tabular}{|c|c|c|c|c|}
\hline
 \multicolumn{2}{|c|}{User-group $m$} &$1$&$2$&$3$ \\
   \hline
   \multicolumn{2}{|c|}{Requested files} &$S_{1}$,~$S_{2}$,~$S_{3}$&$S_{2}$,~$S_{3}$&$S_{1}$,~$S_{4}$ \\
   \hline
   \multicolumn{2}{|c|}{\multirow{4}{*}{Cached Packets (Type-\MyRoman{2})}}&$ S_{1,(1,4)}^{(1)}\oplus S_{4,(1,4)}^{(1)}$ &$S_{1, (1,2)}^{(2)}\oplus S_{2,(1,2)}^{(2)}$&$S_{1,(1,2)}^{(3)}\oplus S_{2,(1,2)}^{(3)}$\\
    \multicolumn{2}{|c|}{ }&$S_{2, (2,4)}^{(1)}\oplus S_{4,(2,4)}^{(1)}$&$S_{1,(1,3)}^{(2)}\oplus S_{3,(1,3)}^{(2)}$&$S_{1,(1,3)}^{(3)}\oplus S_{3,(1,3)}^{(3)}$\\
  \multicolumn{2}{|c|}{ }&$S_{3,(3,4)}^{(1)}\oplus S_{4,(3,4)}^{(1)}$&$S_{2,(2,4)}^{(2)}\oplus S_{4,(2,4)}^{(2)}$&$S_{2,(2,4)}^{(3)}\oplus S_{4,(2,4)}^{(3)}$\\
  \multicolumn{2}{|c|}{ }& &$S_{3,(3,4)}^{(2)}\oplus S_{4,(3,4)}^{(2)}$&$S_{3,(3,4)}^{(3)}\oplus S_{4,(3,4)}^{(3)}$\\
    \hline
 \multirow{6}{*}{Step 1}&\multirow{2}{*}{Transmissions}&$S_{4,(1,4)}^{(1)}$, $S_{4,(2,4)}^{(1)}$& $S_{1,(1,2)}^{(2)}$,~$S_{1,(1,3)}^{(2)}$
 &$S_{2,(1,2)}^{(3)}$,~$S_{3,(1,3)}^{(3)}$ \\
 & &$S_{4,(3,4)}^{(1)}$&$S_{4,(2,4)}^{(2)}$,~$S_{4,(3,4)}^{(2)}$ &$S_{2,(2,4)}^{(3)}$,~$S_{3,(3,4)}^{(3)}$  \\
 \cline{2-5}
&\multirow{4}{*}{Acquisition}&$S_{1,(1,4)}^{(1)}$,~$S_{2,(2,4)}^{(1)}$,~$S_{3,(3,4)}^{(1)}$&$S_{2,(1,2)}^{(2)}$,~$S_{3,(1,3)}^{(2)}$
&$S_{1,(1,2)}^{(3)}$,~$S_{4,(2,4)}^{(3)}$,~$S_{4,(3,4)}^{(3)}$\\
& &$S_{1,(1,2)}^{(2)}$,~$S_{2,(1,2)}^{(3)}$,~$S_{3,(1,3)}^{(3)}$& $S_{2,(2,4)}^{(2)}$,~$S_{3,(3,4)}^{(2)}$&$S_{1,(1,3)}^{(3)}$, $S_{4,(1,4)}^{(1)}$, $S_{4,(2,4)}^{(1)}$\\
& &$S_{1,(1,3)}^{(2)}$,~$S_{2,(2,4)}^{(3)}$,~$S_{3,(3,4)}^{(3)}$&$S_{2,(1,2)}^{(3)}$,~$S_{3,(1,3)}^{(3)}$ &$S_{1,(1,2)}^{(2)}$,~$S_{4,(3,4)}^{(1)}$,~$S_{4,(2,4)}^{(2)}$ \\
& & &$S_{2,(2,4)}^{(3)}$,~$S_{3,(3,4)}^{(3)}$&$S_{1,(1,3)}^{(2)}$,~$S_{4,(3,4)}^{(2)}$,~~~~~~~~~~~\\
\hline
 \multirow{2}{*}{Step 2}&\multirow{1}{*}{Transmissions}& \multicolumn{3}{|c|}{$S_{1,(1,4)}^{(1)}\oplus S_{1,(1,2)}^{(3)}$,~$S_{2,(2,4)}^{(1)}\oplus S_{2,(1,2)}^{(2)}$,~$S_{3,(3,4)}^{(1)}\oplus S_{3,(1,3)}^{(2)}$}\\
\cline{2-5}
&Acquisition &$ S_{1,(1,2)}^{(3)}$,~$S_{2,(1,2)}^{(2)}$,~$S_{3,(1,3)}^{(2)}$ &$S_{2,(2,4)}^{(1)}$,~$S_{3,(3,4)}^{(1)}$
 &$S_{1,(1,4)}^{(1)}$\\
\hline
\multicolumn{2}{|c|}{Remaining fragments}&$\left\backslash\vphantom{\frac{a^{1}}{b^{2}}}\right.$&$S_{2,(2,4)}^{(2)}$,~$S_{3,(3,4)}^{(2)}$&$S_{1,(1,3)}^{(3)}$\\
\hline
\end{tabular}\caption{Illustration for the delivery of the fragments in Type-\MyRoman{2} packets.}\label{tab:DeliverySchemeofPacketType2}
\end{table*}
\begin{example} Assume $(N,M,D,\alpha)=(4,3,7,2)$, where $N_{\mathrm{R}}=N=4$, $\mathcal{D}_{1}=\{1,2,3\}$, $\mathcal{D}_{2}=\{2,3\}$ and $\mathcal{D}_{3}=\{1,4\}$. Then all Type-\MyRoman{2} packets in the three caches are given in the third row of Table \ref{tab:DeliverySchemeofPacketType2}. Since each packet contains only one local fragment, the server first transmits the other $\alpha-1=1$ fragment. Taking packet $S_{1,(1,4)}^{(1)}\oplus S_{4,(1,4)}^{(1)}$ in cache 1 for example, the server transmits $S_{4,(1,4)}^{(1)}$, then user-group 1 can obtain $S_{1,(1,4)}^{(1)}$ by XOR decoding and meanwhile other user-groups can obtain $S_{4,(1,4)}^{(1)}$ from the direct transmission of the server. Note that $S_{1,(1,4)}^{(1)}$ is also requested by user-group 3, whose cache also has local fragments from $S_{1}$: $S_{1,(1,2)}^{(3)}$ and $S_{1,(1,3)}^{(3)}$. To deliver these local fragments for the two user-groups, the server can then transmit a pairwise-coded packet encoded by XORing $S_{1,(1,4)}^{(1)}$ together with $S_{1,(1,2)}^{(3)}$ or $S_{1,(1,3)}^{(3)}$. The rule to form pairwise-coded packets is that no local fragment is repeatedly XORed between any two caches. In our example we choose to transmit $S_{1,(1,4)}^{(1)}\oplus S_{1,(1,2)}^{(3)}$, and thus $S_{1,(1,3)}^{(3)}$ is a remaining unpaired fragment which will be delivered at the last stage. In such a way, the local fragments that can be pairwise-coded are delivered to multiple user-groups (see Step 2 in the table) and the untransmitted local fragments requested by multiple user-groups are shown in the last row.
\end{example}
 \par Next, we present the general delivery scheme according to Type-\MyRoman{2} packets, which is divided into the following two steps:
\subsubsection{Step 1} Transmit $\alpha-1$ fragments encoded in each packet $P_{(n_{1},...,n_{\alpha})}^{(m)}$ except the local fragment $S_{n_{i},(n_{1},...,n_{\alpha})}^{(m)}$, which are
  \begin{equation}\label{equ:TransmittedContents_Type2_a}
 S_{n,(n_{1},...,n_{\alpha})}^{(m)},~ n=n_{1},...,n_{i-1},n_{i+1},...,n_{\alpha}.
\end{equation}
Then user-group $m$ can obtain $S_{n_{i},(n_{1},...,n_{\alpha})}^{(m)}$ by XOR decoding since $n_{i}\in\mathcal{D}_{m}$, and meanwhile other user-groups in $\mathcal{M}\setminus\{m\}$ can obtain their requested fragments given in (\ref{equ:TransmittedContents_Type2_a}) from the direct transmissions.
\par For any $n_{i}\in\mathcal{D}_{m}$, the number of Type-\MyRoman{2} packets is $\textit{\textsf{C}}_{N_{\mathrm{R}}-D_{m} }^{\alpha-1}$ and thus the number of transmissions is $\textit{\textsf{C}}_{N_{\mathrm{R}}-D_{m} }^{\alpha-1}(\alpha-1)$. Since the distinct request number of user-group $m$ is $D_{m}$, the number of transmissions for user-group $m$ is $D_{m}\textit{\textsf{C}}_{N_{\mathrm{R}}-D_{m} }^{\alpha-1}(\alpha-1)$.
 Then the delivery load for the $M$ user-groups is given by
    \begin{equation}\label{equ:NumofBroadcastforType2_HMS_sub1}
     T_{\mathrm{\MyRoman{2}}}^{(1)}= \sum_{m=1}^{M}\textsf{\textit{C}}_{N_{\mathrm{R}}-D_{m} }^{\alpha-1}D_{m} (\alpha-1).
    \end{equation}
\subsubsection{Step 2} Deliver the local fragments of $S_{n_{i}}$ for any $n_{i}\in \mathcal{N}_{\mathrm{R}}$ that can be pairwise-encoded among the caches of $\mathcal{M}(n_{i})$ , where $\mathcal{M}(n_{i})$ denotes the set of user-groups requesting $S_{n_{i}}$. Given any $m\in\mathcal{M}(n_{i})$, the number of local fragments of $S_{n_{i}}$ in cache $m$ is $\textit{\textsf{C}}_{N_{\mathrm{R}}-D_{m}}^{\alpha-1}$. Define $\overline{m}\triangleq\mathop{\arg\min}\limits_{m\in\mathcal{M}(n_{i})}\textit{\textsf{C}}_{N_{\mathrm{R}}-D_{m}}^{\alpha-1}$.
Then cache $\overline{m}$ has the minimum number of local fragments of $S_{n_{i}}$, given by
 \begin{equation}\label{equ:TransmittedContents_Type2_b0}
   S_{n_{i},(n_{1},...,n_{\alpha})}^{(\overline{m})},~ (n_{1},...,n_{\alpha})\in\widetilde{\mathcal{A}}~\mathrm{and}~\{n_{1},...,n_{\alpha}\} \cap \mathcal{D}_{\overline{m}}=\{n_{i}\}.
 \end{equation}
 Note that the local fragments of $S_{n_{i}}$ in different caches of $\mathcal{M}(n_{i})$ may come from the packets with different file combinations, we denote them in cache $m\in\mathcal{M}(n_{i})\setminus\{\overline{m}\}$ by
 \begin{equation}\label{equ:TransmittedContents_Type2_b1}
   S_{n_{i},(n_{1}^{(m)},...,n_{\alpha}^{(m)})}^{(m)},~ (n_{1}^{(m)},...,n_{\alpha}^{(m)})\in\widetilde{\mathcal{A}}~\mathrm{and}~\{n_{1}^{(m)},...,n_{\alpha}^{(m)}\} \cap \mathcal{D}_{m}=\{n_{i}\},
 \end{equation}
 where $(n_{1}^{(m)},...,n_{\alpha}^{(m)})\neq(n_{1},...,n_{\alpha})$ may hold.
Take each local fragment of $S_{n_{i}}$ in cache $\overline{m}$ as a reference fragment, and XOR it together with one local fragment of $S_{n_{i}}$ from any other cache $m\in\mathcal{M}(n_{i})\setminus\{\overline{m}\}$ to form $|\mathcal{M}(n_{i})|-1$ pairwise-coded packets under the condition that no local fragment is repeatedly XORed with different reference fragments, given by
  \begin{equation}\label{equ:TransmittedContents_Type2_b}
   S_{n_{i},(n_{1},...,n_{\alpha})}^{(\overline{m})} \oplus S_{n_{i},(n_{1}^{(m)},...,n_{\alpha}^{(m)})}^{(m)},~ m\in\mathcal{M}(n_{i})\setminus\{\overline{m}\}.
\end{equation}
The above packets are transmitted and then user-group $\overline{m}$ can obtain $S_{n_{i},(n_{1}^{(m)},...,n_{\alpha}^{(m)})}^{(m)},~ m\in\mathcal{M}(n_{i})\setminus\{\overline{m}\}$ in them via XOR operations since it caches $S_{n_{i},(n_{1},...,n_{\alpha})}^{(\overline{m})}$; and any other user-group $m\in\mathcal{M}(n_{i})\setminus\{\overline{m}\}$ can obtain $S_{n_{i},(n_{1},...,n_{\alpha})}^{(\overline{m})}$ and $S_{n_{i},(n_{1}^{(k)},...,n_{\alpha}^{(k)})}^{(k)},~ k\in\mathcal{M}(n_{i})\setminus\{\overline{m},m\}$ in them via XOR operations since it caches $S_{n_{i},(n_{1}^{(m)},...,n_{\alpha}^{(m)})}^{(m)}$. Thus for the transmissions of (\ref{equ:TransmittedContents_Type2_b}) for each reference fragment $S_{n_{i},(n_{1},...,n_{\alpha})}^{(\overline{m})}$, every user-group of $\mathcal{M}(n_{i})$ can obtain the $|\mathcal{M}(n_{i})|$ local fragments of $S_{n_{i}}$ encoded in them, each from one cache of $\mathcal{M}(n_{i})$.
\par As there are $\textit{\textsf{C}}_{N_{\mathrm{R}}-D_{\overline{m}}}^{\alpha-1}$ local fragments of $S_{n_{i}}$ in cache $\overline{m}$, the number of transmitted packets for any $n_{i}\in\mathcal{N}_{\mathrm{R}}$ is $\left(|\mathcal{M}(n_{i})|-1\right) \textit{\textsf{C}}_{N_{\mathrm{R}}-D_{\overline{m}}}^{\alpha-1}$. Since there are $N_{\mathrm{R}}$ requested files, the total number of transmitted packets is
 \begin{equation}\label{equ:NumofBroadcastforType2_HMS_sub2}
 \begin{split}
 T_{\mathrm{\MyRoman{2}}}^{(2)}&=\sum\limits_{n_{i}\in \mathcal{N}_{\mathrm{R}}}\left(|\mathcal{M}(n_{i})|-1\right)\textit{\textsf{C}}_{N_{\mathrm{R}}-D_{\overline{m}}}^{\alpha-1} =\sum\limits_{n_{i}\in \mathcal{N}_{\mathrm{R}}} \left(|\mathcal{M}(n_{i})|-1\right)\min\limits_{m\in\mathcal{M}(n_{i})}\textit{\textsf{C}}_{N_{\mathrm{R}}-D_{m}}^{\alpha-1}.
\end{split}
\end{equation}
Note that cache $m\in\mathcal{M}(n_{i})$ contains $\textit{\textsf{C}}_{N_{\mathrm{R}}-D_{m}}^{\alpha-1}$ local fragments of $S_{n_{i}}$, and $\textit{\textsf{C}}_{N_{\mathrm{R}}-D_{\overline{m}}}^{\alpha-1}$ of them have been pairwise-encoded and transmitted. After the transmissions of (\ref{equ:TransmittedContents_Type2_b}) for each $S_{n_{i},(n_{1},...,n_{\alpha})}^{(\overline{m})}$, the number of remaining unpaired local fragments for all $n_{i}\in\mathcal{N}_{\mathrm{R}}$ that are requested by multiple user-groups but untransmitted is
\begin{equation}\label{equ:RemaininguntransmittedFragments_Type2}
  T_{\mathrm{\MyRoman{2}}}^{(\mathrm{RM})}
  =\sum\limits_{n_{i}\in \mathcal{N}_{\mathrm{R}}} \sum\limits_{m\in\mathcal{M}(n_{i})}\left(\textit{\textsf{C}}_{N_{\mathrm{R}}-D_{m}}^{\alpha-1}- \min\limits_{m\in\mathcal{M}(n_{i})}\textit{\textsf{C}}_{N_{\mathrm{R}}-D_{m}}^{\alpha-1}\right).
\end{equation}
These remaining fragments will be delivered at the last stage, as will be discussed in Section \ref{sec:subsec:DeliveryLsatStage}.
\par Then based on the above analysis, the delivery load of Type-\MyRoman{2} packets by combining (\ref{equ:NumofBroadcastforType2_HMS_sub1}) with (\ref{equ:NumofBroadcastforType2_HMS_sub2}) is given by
   \begin{equation}\label{equ:NumofBroadcastforType2_HCMS}
  T_{\mathrm{\MyRoman{2}}}=T_{\mathrm{\MyRoman{2}}}^{(1)}+T_{\mathrm{\MyRoman{2}}}^{(2)}
  =\sum_{m=1}^{M}\textsf{\textit{C}}_{N_{\mathrm{R}}-D_{m} }^{\alpha-1}D_{m}(\alpha-1) +\sum\limits_{n_{i}\in \mathcal{N}_{\mathrm{R}}} \left(|\mathcal{M}(n_{i})|-1\right)\min\limits_{m\in\mathcal{M}(n_{i})}\textit{\textsf{C}}_{N_{\mathrm{R}}-D_{m}}^{\alpha-1}.
\end{equation}
\subsection{Type-\emph{\MyRoman{3}} Cached Packets}
A cached packet $P_{(n_{1},...,n_{\alpha})}^{(m)}$ is Type-\MyRoman{3} if it is encoded by requested fragments only, among which there are more than one local fragment. Such packet can be characterized by
 \begin{equation}\label{equ:Packets_Type3}
    P_{(n_{1},...,n_{\alpha})}^{(m)}=\bigoplus\limits_{i=1}^{\alpha} S_{n_{i},(n_{1},...,n_{\alpha})}^{(m)},~ (n_{1},...,n_{\alpha})\in\widetilde{\mathcal{A}}~\mathrm{and}~|\{n_{1},...,n_{\alpha}\} \cap \mathcal{D}_{m}|>1,
 \end{equation}
 where $\widetilde{\mathcal{A}}\subseteq \mathcal{N}_{\mathrm{R}}$ is defined in (\ref{equ:CodedPrefetching:Definition3}). We first use the following example to illustrate Type-\MyRoman{3} packets and the delivery scheme according to them.
  \begin{table*} [ht]
  \centering
\begin{tabular}{|c|c|c|c|c|}
\hline
 \multicolumn{2}{|c|}{User-group $m$} &$1$&$2$&$3$ \\
   \hline
   \multicolumn{2}{|c|}{Requested files} &$S_{1}$,~$S_{2}$,~$S_{3}$&$S_{2}$,~$S_{3}$,~$S_{4}$&$S_{2}$,~$S_{3}$,~$S_{5}$ \\
   \hline
   \multicolumn{2}{|c|}{\multirow{3}{*}{Cached Packets (Type-\MyRoman{3})}}&$S_{1,(1,2)}^{(1)}\oplus S_{2,(1,2)}^{(1)}$&$S_{2,(2,3)}^{(2)}\oplus S_{3,(2,3)}^{(2)}$&$S_{2,(2,3)}^{(3)}\oplus S_{3,(2,3)}^{(3)}$\\
   \multicolumn{2}{|c|}{ }&$S_{1,(1,3)}^{(1)}\oplus S_{3,(1,3)}^{(1)}$&$S_{2,(2,4)}^{(2)}\oplus S_{4,(2,4)}^{(2)}$&$S_{2,(2,5)}^{(3)}\oplus S_{5,(2,5)}^{(3)}$\\
   \multicolumn{2}{|c|}{ }&$S_{2,(2,3)}^{(1)}\oplus S_{3,(2,3)}^{(1)}$&$S_{3,(3,4)}^{(2)}\oplus S_{4,(3,4)}^{(2)}$&$S_{3,(3,5)}^{(3)}\oplus S_{5,(3,5)}^{(3)}$\\
    \hline
  \multirow{4}{*}{Step 1}&Transmissions&$S_{2,(1,2)}^{(1)}$, $S_{3,(1,3)}^{(1)}$, $S_{3,(2,3)}^{(1)}$& $S_{2,(2,3)}^{(2)}$,~$S_{2,(2,4)}^{(2)}$,~$S_{3,(3,4)}^{(2)}$&$S_{2,(2,3)}^{(3)}$, $S_{2,(2,5)}^{(3)}$, $S_{3,(3,5)}^{(3)}$\\
 \cline{2-5}
 &\multirow{4}{*}{Acquisition}& $S_{1,(1,2)}^{(1)}$, $S_{2,(1,2)}^{(1)}$, $S_{3,(1,3)}^{(1)}$&$S_{2,(2,3)}^{(2)}$,~$S_{3,(2,3)}^{(2)}$,~$S_{4,(2,4)}^{(2)}$&$S_{2,(2,3)}^{(3)}$, $S_{3,(2,3)}^{(3)}$, $S_{5,(2,5)}^{(3)}$ \\
& &$S_{1,(1,3)}^{(1)}$, $S_{2,(2,3)}^{(1)}$, $S_{3,(2,3)}^{(1)}$&$S_{2,(2,4)}^{(2)}$,~$S_{3,(3,4)}^{(2)}$,~$S_{4,(3,4)}^{(2)}$&$S_{2,(2,5)}^{(3)}$, $S_{3,(3,5)}^{(3)}$, $S_{5,(3,5)}^{(3)}$ \\
 & &$S_{2,(2,3)}^{(2)}$,~$S_{2,(2,4)}^{(2)}$,~$S_{3,(3,4)}^{(2)}$&$S_{2,(1,2)}^{(1)}$, $S_{3,(1,3)}^{(1)}$, $S_{3,(2,3)}^{(1)}$&$S_{2,(1,2)}^{(1)}$, $S_{3,(1,3)}^{(1)}$, $S_{3,(2,3)}^{(1)}$ \\
  & &$S_{2,(2,3)}^{(3)}$, $S_{2,(2,5)}^{(3)}$, $S_{3,(3,5)}^{(3)}$&$S_{2,(2,3)}^{(3)}$, $S_{2,(2,5)}^{(3)}$, $S_{3,(3,5)}^{(3)}$&$S_{2,(2,3)}^{(2)}$,~$S_{2,(2,4)}^{(2)}$,~$S_{3,(3,4)}^{(2)}$\\
 \hline
  \multirow{2}{*}{Step 2}&Transmissions& \multicolumn{3}{c|}{$S_{2,(2,3)}^{(1)}\oplus S_{3,(2,3)}^{(2)}$, $S_{2,(2,3)}^{(1)}\oplus S_{3,(2,3)}^{(3)}$} \\
 \cline{2-5}
 & Acquisition& $S_{3,(2,3)}^{(2)}$, $S_{3,(2,3)}^{(3)}$&$S_{2,(2,3)}^{(1)}$, $S_{3,(2,3)}^{(3)}$&$S_{2,(2,3)}^{(1)}$, $S_{3,(2,3)}^{(2)}$ \\
 \hline
 \multicolumn{2}{|c|}{Remaining fragments}&$\left\backslash\vphantom{\frac{a^{1}}{b^{2}}}\right.$&$\left\backslash\vphantom{\frac{a^{1}}{b^{2}}}\right.$&$\left\backslash\vphantom{\frac{a^{1}}{b^{2}}}\right.$ \\
 \hline
\end{tabular}\caption{Illustration for the delivery of the fragments in Type-\MyRoman{3} packets.}\label{tab:DeliverySchemeofPacketType3}
\end{table*}
 \begin{example}
 Assume $(N,M,D,\alpha)=(5,3,9,2)$, where $N_{\mathrm{R}}=N=5$, $\mathcal{D}_{1}=\{1,2,3\}$, $\mathcal{D}_{2}=\{2,3,4\}$ and $\mathcal{D}_{3}=\{2,3,5\}$. Then all the Type-\MyRoman{3} packets in the three caches are given in the third row of Table \ref{tab:DeliverySchemeofPacketType3}. Similar to the delivery scheme according to Type-\MyRoman{2} packets, the server first needs to transmit $\alpha-1=1$ fragment for each packet, by not transmitting the local fragment that is least requested by other user-groups. Taking the packets in cache 1 for example, the untransmitted local fragments for $S_{1,(1,2)}^{(1)}\oplus S_{2,(1,2)}^{(1)}$ and $S_{1,(1,3)}^{(1)}\oplus S_{3,(1,3)}^{(1)}$ are $S_{1,(1,2)}^{(1)}$ and $S_{1,(1,3)}^{(1)}$, respectively; while the untransmitted local fragment for $S_{2,(2,3)}^{(1)}\oplus S_{3,(2,3)}^{(1)}$ is $S_{2,(2,3)}^{(1)}$ or $ S_{3,(2,3)}^{(1)}$ since the numbers of user-groups requesting $S_{2}$ and $S_{3}$ are the same. As $S_{1}$ is only requested by user-group 1, no more transmission is needed for $S_{1,(1,2)}^{(1)}$ and $S_{1,(1,3)}^{(1)}$ after Step 1 but more transmissions are needed for $S_{2,(2,3)}^{(1)}$ as it is also requested by other user-groups. Similarly, after Step 1 the untransmitted local fragments requested by multiple user-groups in caches 2 and 3 are $S_{3,(2,3)}^{(2)}$ and $ S_{3,(2,3)}^{(3)}$, respectively. Take $S_{2,(2,3)}^{(1)}$ as the reference fragment and XOR it together with $S_{3,(2,3)}^{(2)}$ and $ S_{3,(2,3)}^{(3)}$, respectively, to form two packets, and then transmit them. Thus the three untransmitted local fragments can be obtained by the three user-groups. After it, all local fragments requested by multiple user-groups are obtained.
\end{example}
 \par Next, we present the general delivery scheme according to Type-\MyRoman{3} packets, which can be implemented by the following two steps:
\subsubsection{Step 1} Transmit $\alpha-1$ fragments encoded in each packet $P_{(n_{1},...,n_{\alpha})}^{(m)}$ except the local fragment that is least requested by other user-groups denoted as $S_{n_{i_{0}},(n_{1},...,n_{\alpha})}^{(m)}$. Thus for any $n_{i}\in\{n_{1},...,n_{\alpha}\}\cap\mathcal{D}_{m}$ we have $|\mathcal{M}(n_{i_{0}})|\leq|\mathcal{M}(n_{i})|$, and the transmitted $\alpha-1$ fragments are
          \begin{equation}\label{equ:TransmittedContents_Type3_a}
 S_{n,(n_{1},...,n_{\alpha})}^{(m)},~ n=n_{1},...,n_{i_{0}-1},n_{i_{0}+1},...,n_{\alpha}.
\end{equation}
\par As the number of the distinct requests of user-group $m$ is $D_{m}$, the number of Type-\MyRoman{3} packets stored in cache $m$ is
\begin{equation}\label{equ:NumType3foreach_HCMS}
\sum\limits_{n=2}^{\min\{D_{m},\alpha\}}  \textit{\textsf{C}}_{D_{m}}^{n} \textit{\textsf{C}}_{N_{\mathrm{R}}-D_{m}}^{\alpha-n} .
\end{equation}
Since each packet needs transmitting $\alpha-1$ fragments, the delivery load for the $M$ user-groups in Step 1 is given by
    \begin{equation}\label{equ:NumofBroadcastforType2a_HMS_sub1}
     T_{\mathrm{\MyRoman{3}}}^{(1)}= \sum_{m=1}^{M}\sum\limits_{n=2}^{\min\{D_{m},\alpha\}}\textit{\textsf{C}}_{D_{m}}^{n} \textit{\textsf{C}}_{N_{\mathrm{R}}-D_{m}}^{\alpha-n} (\alpha-1).
    \end{equation}
\subsubsection{Step 2} Deliver the untransmitted local fragments $\{S_{n_{i_{0}},(n_{1},...,n_{\alpha})\in\widetilde{\mathcal{A}}}^{(m\in\mathcal{M})}:n_{i_{0}}\in \{n_{1},...,n_{\alpha}\}\cap\mathcal{D}_{m}\}$ that are requested by multiple user-groups and can be pairwise-encoded among the caches of $\mathcal{M}$. Letting $\sigma_m$ denote the number of  distinct files requested by user-group $m$ only, the number of untransmitted local fragments that are requested by multple user-groups in cache $m$ is
     \begin{equation}\label{equ:NumofBroadcastforType2a_HMS_Process_sub3}
    L_{m}^{(\mathrm{UnTr})}=\sum\limits_{n=2}^{\min\{D_{m},\alpha\}}\textit{\textsf{C}}_{D_{m}-\sigma_{m}}^{n} \textit{\textsf{C}}_{N_{\mathrm{R}}-D_{m}}^{\alpha-n} .
    \end{equation}
Denote $\widetilde{m}\triangleq\mathop{\arg\min}\limits_{m\in\mathcal{M}} L_{m}^{(\mathrm{UnTr})}$ and take cache $\widetilde{m}$ as the reference cache. Then for each multi-requested local fragment $S_{n_{i_{0}},(n_{1},...,n_{\alpha})}^{(\widetilde{m})}$ in cache $\widetilde{m}$, a group of $M$ untransmitted local fragments that are requested by multiple user-groups and unrepeatedly selected from the $M$ caches can be picked out, which are denoted as
   \begin{equation}\label{equ:TransmittedContents_Type3_b0}
   S_{n_{i_{0}},(n_{1},...,n_{\alpha})}^{(\widetilde{m})}~ \mathrm{and}~S_{n_{i_{0}}^{(m)},(n_{1}^{(m)},...,n_{\alpha}^{(m)})}^{(m)},~m=1,...,\widetilde{m}-1,\widetilde{m}+1,...,M,
 \end{equation}
 where $n_{i_{0}}^{(m)}\neq n_{i_{0}}$ or $(n_{1}^{(m)},...,n_{\alpha}^{(m)})\neq(n_{1},...,n_{\alpha})$ may hold as the untransmitted local fragments in different caches may come from different files and also from different file combinations. For each group of the $M$ untransmitted local fragments, XOR one fragment from any cache $m\in\mathcal{M}\setminus\{\widetilde{m}\}$ with that from cache $\widetilde{m}$ to form $M-1$ pairwise-coded packets to obtain
   \begin{equation}\label{equ:TransmittedContents_Type3_b}
     S_{n_{i_{0}},(n_{1},...,n_{\alpha})}^{(\widetilde{m})}\oplus S_{n_{i_{0}}^{(m)},(n_{1}^{(m)},...,n_{\alpha}^{(m)})}^{(m)},~m=1,...,\widetilde{m}-1,\widetilde{m}+1,...,M.
 \end{equation}
The above packets are transmitted. Similar to the case of Type-\MyRoman{2} packets, then every user-group can obtain its requested fragments in the transmitted packets via XOR operations since it caches one of the requested fragments. Thus the untransmitted local fragments that are pairwise-encoded can be delivered to the requesting user-groups.
\par Note that cache $\widetilde{m}$ has $L_{\widetilde{m}}^{(\mathrm{UnTr})}$ multi-requested local fragments $\{S_{n_{i_{0}},(n_{1},...,n_{\alpha})\in\widetilde{\mathcal{A}}}^{(\widetilde{m})}:n_{i_{0}}\in \{n_{1},...,n_{\alpha}\}\cap\mathcal{D}_{\widetilde{m}}\}$, each corresponding to a group of $M$ untransmitted local fragments from the $M$ caches and leading to the transmissions of $M-1$ pairwise-coded packets. Thus the total number of transmitted packets is
 \begin{equation}\label{equ:NumofBroadcastforType2a_HMS_sub3}
     \begin{split}
     T_{\mathrm{\MyRoman{3}}}^{(2)}=&(M-1)L_{\widetilde{m}}^{(\mathrm{UnTr})}=(M-1)\min\limits_{m\in\mathcal{M}}L_{m}^{(\mathrm{UnTr})};
      \end{split}
   \end{equation}
and the number of remaining untransmitted local fragments that are requested by multiple user-groups is given by
    \begin{equation}\label{equ:RemaininguntransmittedFragments_Type3}
      T_{\mathrm{\MyRoman{3}}}^{(\mathrm{RM})}=\sum\limits_{m=1}^{M}L_{m}^{(\mathrm{UnTr})} -ML_{\widetilde{m}}^{(\mathrm{UnTr})}=\sum\limits_{m=1}^{M}L_{m}^{(\mathrm{UnTr})} -M\min\limits_{m\in\mathcal{M}}L_{m}^{(\mathrm{UnTr})}.
    \end{equation}
    Similarly, the delivery of these remaining fragments will be conducted at the last stage.
   \par Then combining (\ref{equ:NumofBroadcastforType2a_HMS_sub1}) with (\ref{equ:NumofBroadcastforType2a_HMS_sub3}) we can obtain the delivery load given by
\begin{equation}\label{equ:NumofBroadcastforType2a_HCMS}
\begin{split}
  T_{\mathrm{\MyRoman{3}}}&=T_{\mathrm{\MyRoman{3}}}^{(1)}+T_{\mathrm{\MyRoman{3}}}^{(2)}=\sum_{m=1}^{M}\sum\limits_{n=2}^{\min\{D_{m},\alpha\}}\textit{\textsf{C}}_{D_{m}}^{n} \textit{\textsf{C}}_{N_{\mathrm{R}}-D_{m}}^{\alpha-n}(\alpha-1) +(M-1)\min\limits_{m\in\mathcal{M}}L_{m}^{(\mathrm{UnTr})},
  \end{split}
\end{equation}
where $L_{m}^{(\mathrm{UnTr})}$ is given in (\ref{equ:NumofBroadcastforType2a_HMS_Process_sub3}).
\subsection{Type-\emph{\MyRoman{4}} Cached Packets}
A cached packet $P_{(n_{1},...,n_{\alpha})}^{(m)}$ is Type-\MyRoman{4} if it is encoded by requested fragments only, all of which are not local fragments. Such packet can be characterized by
 \begin{equation}\label{equ:Packets_Type4}
    P_{(n_{1},...,n_{\alpha})}^{(m)}=\bigoplus\limits_{i=1}^{\alpha} S_{n_{i},(n_{1},...,n_{\alpha})}^{(m)},~ (n_{1},...,n_{\alpha})\in\widetilde{\mathcal{A}}~\mathrm{and}~\{n_{1},...,n_{\alpha}\} \cap \mathcal{D}_{m}=\emptyset,
 \end{equation}
 where $\widetilde{\mathcal{A}}\subseteq \mathcal{N}_{\mathrm{R}}$ is defined in (\ref{equ:CodedPrefetching:Definition3}). Define the \emph{packet-group} as a group of cached packets each from a different cache and define $\mathcal{V}\triangleq\{r_{0},r_{1},...,r_{\alpha}\}\subseteq \mathcal{N}_{\mathrm{R}}$ as an $(\alpha+1)$-request set such that no user-group requests two or more than two of its elements, i.e.,
    \begin{equation}\label{equ:PacketofType4_ConsiditionforStep1_a}
 \mathcal{M}(r_{i})\cap\mathcal{M}(r_{j})=\emptyset,~\forall r_{i}\neq r_{j}\in \mathcal{V},
  \end{equation}
where ${\mathcal{M}}(r)$ denotes the set of all the user-groups in $\mathcal{M}$ requesting file $S_{r}$. Letting $\mathcal{M}_{\mathcal{V}}\triangleq\bigcup\limits_{r\in\mathcal{V}}\mathcal{M}(r)$, we have $|\mathcal{D}_{m}\cap\mathcal{V}|=1$ for any $m\in\mathcal{M}_{\mathcal{V}}$ and there exists a packet-group consisting of $M_{\mathcal{V}}=|\mathcal{M}_{\mathcal{V}}|$ Type-\MyRoman{4} packets $\{P_{(n_{1}^{(m)},...,n_{\alpha}^{(m)})}^{(m)}:m\in\mathcal{M}_{\mathcal{V}}\}$ such that each packet is encoded by $\alpha$ fragments from $\alpha$ files in $\mathcal{V}$, i.e., $\{n_{1}^{(m)},...,n_{\alpha}^{(m)}\}\subset \mathcal{V}$ for any $m\in\mathcal{M}_{\mathcal{V}}$, according to the definition of Type-\MyRoman{4} packets. Then in the delivery of Type-\MyRoman{4} packets, the server first delivers the fragments in each such packet-group corresponding to an $(\alpha+1)$-request set $\mathcal{V}$ and then delivers the fragments in the remaining Type-\MyRoman{4} packets, which result in three steps in the delivery according to Type-\MyRoman{4} packets. To illustrate such type of packets and the delivery scheme according to it, we first present the following example.
 \begin{table*} [ht]
  \centering
  \renewcommand\arraystretch{1.3}
\begin{tabular}{|c|c|c|c|c|}
\hline
  \multicolumn{2}{|c|}{User-group $m$} &$1$&$2$&$3$  \\
   \hline
   \multicolumn{2}{|c|}{Requested files} &$S_{1}$, $S_{2}$, $S_{4}$& $S_{2}$, $S_{3}$&$S_{4}$, $S_{5}$\\
   \hline
  \multicolumn{2}{|c|}{\multirow{3}{*}{Cached Packets (Type-\MyRoman{4})}}&\fbox{$S_{3,(3,5)}^{(1)}$}$\oplus S_{5,(3,5)}^{(1)}$&$S_{1,(1,5)}^{(2)}\oplus$\fbox{$ S_{5,(1,5)}^{(2)}$}&\fbox{$S_{1,(1,3)}^{(3)}$}$\oplus S_{3,(1,3)}^{(3)}$ \\
 \multicolumn{2}{|c|}{ }&&$S_{4,(4,5)}^{(2)}\oplus \overline{S_{5,(4,5)}^{(2)}}$&$S_{2,(2,3)}^{(3)}\oplus \overline{S_{3,(2,3)}^{(3)}}$\\
  \multicolumn{2}{|c|}{ }&&$S_{1,(1,4)}^{(2)}\oplus S_{4,(1,4)}^{(2)}$&$S_{1,(1,2)}^{(3)}\oplus S_{2,(1,2)}^{(3)}$ \\
   \hline
   \multirow{3}{*}{Step 1}&\multirow{2}{*}{Transmissions }&$S_{5,(3,5)}^{(1)}\oplus$\fbox{$ S_{5,(1,5)}^{(2)}$}&$S_{1,(1,5)}^{(2)}\oplus$\fbox{$S_{1,(1,3)}^{(3)}$}& $S_{3,(1,3)}^{(3)}\oplus$\fbox{$S_{3,(3,5)}^{(1)}$}\\
  \cline{3-5}
  & &\multicolumn{3}{c|}{\fbox{$S_{1,(1,3)}^{(3)}$}$\oplus$ \fbox{$S_{3,(3,5)}^{(1)}$}$\oplus$\fbox{$ S_{5,(1,5)}^{(2)}$}}\\
   \cline{2-5}
   &Acquisition&$S_{1,(1,5)}^{(2)}$, \fbox{$S_{1,(1,3)}^{(3)}$}&$S_{3,(1,3)}^{(3)}$, \fbox{$S_{3,(3,5)}^{(1)}$}&$S_{5,(3,5)}^{(1)}$, \fbox{$ S_{5,(1,5)}^{(2)}$}  \\
\hline
 \multirow{2}{*}{Step 2}& Transmissions  &$\left\backslash\vphantom{\frac{a^{\frac{1}{2}}}{b^{\frac{1}{2}}}}\right.$&\multicolumn{2}{c|}{$S_{4,(4,5)}^{(2)}$, $\overline{S_{5,(4,5)}^{(2)}}\oplus \overline{S_{3,(2,3)}^{(3)}}$, $S_{2,(2,3)}^{(3)}$} \\%
\cline{2-5}
& Acquisition &$S_{4,(4,5)}^{(2)}$, $S_{2,(2,3)}^{(3)}$&$\overline{S_{3,(2,3)}^{(3)}}$, $S_{2,(2,3)}^{(3)}$& $S_{4,(4,5)}^{(2)}$, $\overline{S_{5,(4,5)}^{(2)}}$ \\
\hline
\multirow{2}{*}{Step 3 }& Transmissions &$\left\backslash\vphantom{\frac{a^{\frac{1}{2}}}{b^{\frac{1}{2}}}}\right.$&$S_{4,(1,4)}^{(2)}$&$S_{2,(1,2)}^{(3)}$ \\%
\cline{2-5}
 & Acquisition &$S_{4,(1,4)}^{(2)}$, $S_{2,(1,2)}^{(3)}$&$S_{2,(1,2)}^{(3)}$&$S_{4,(1,4)}^{(2)}$ \\
\hline
\multicolumn{2}{|c|}{Remaining fragments}& $\left\backslash\vphantom{\frac{a^{\frac{1}{2}}}{b^{\frac{1}{2}}}}\right.$
&$S_{1,(1,4)}^{(2)}$&$S_{1,(1,2)}^{(3)}$\\
        \hline
\end{tabular}\caption{Illustration for the delivery of the fragments in Type-\MyRoman{4} packets.}\label{tab:DeliverySchemeofPacketType4}
\end{table*}
 \begin{example}\label{exm:ExampleforType4Packets}
 Assume $(N,M,D,\alpha)=(5,3,7,2)$, where $N_{\mathrm{R}}=N=5$, $\mathcal{D}_{1}=\{1,2,4\}$, $\mathcal{D}_{2}=\{2,3\}$ and $\mathcal{D}_{3}=\{4,5\}$. Then all the Type-\MyRoman{4} packets in the three caches are given in the third row of Table \ref{tab:DeliverySchemeofPacketType4}. According to the requests of the network, it can be seen that there exists one $(\alpha+1)$-request set $\mathcal{V}=\{1,3,5\}$ that satisfies (\ref{equ:PacketofType4_ConsiditionforStep1_a}), which leads to $\mathcal{M}_{\mathcal{V}}=\{1,2,3\}$ and the packet-group corresponding to $\mathcal{V}$ given by $\{P_{(3,5)}^{(1)},P_{(1,5)}^{(2)},P_{(1,3)}^{(3)}\}$ and shown in the first row in the cached packets. Then, in Step 1, select $\alpha+1=3$ fragments from the $3$ packets such that each selected fragment boxed in the table corresponds to a distinct request from $\mathcal{V}$. XOR each unselected fragment together with a selected fragment of the same file to form $M_{\mathcal{V}}\alpha-(\alpha+1)=3$ transmitted packets and meanwhile XOR all the $3$ selected fragments together to form one more transmitted packet, as seen from the transmissions in the fourth and fifth rows of the table. Then the delivery of the $M_{\mathcal{V}}\alpha=6$ fragments in the packet-group can be completed. Taking user-group 1 for example, it can first obtain $\boxed{S_{1,(1,3)}^{(3)}}$, which is the selected fragment of $S_{1}$, by XORing its cached packet $\boxed{S_{3,(3,5)}^{(1)}}\oplus S_{5,(3,5)}^{(1)}$ together with the transmitted packets $S_{5,(3,5)}^{(1)}\oplus\boxed{S_{5,(1,5)}^{(2)}}$ and $\boxed{S_{1,(1,3)}^{(3)}}\oplus \boxed{S_{3,(3,5)}^{(1)}}\oplus\boxed{S_{5,(1,5)}^{(2)}}$, then it can obtain $S_{1,(1,5)}^{(2)}$, which is the unselected fragment of $S_{1}$, by XORing $\boxed{S_{1,(1,3)}^{(3)}}$ together with the transmitted packet $S_{1,(1,5)}^{(2)}\oplus\boxed{S_{1,(1,3)}^{(3)}}$. The delivery load for Step 1 is 4 and thus the delivery gain is 2, where the delivery gain denotes the number of reduced transmissions compared with direct transmissions. Over the remaining Type-\MyRoman{4} packets, there is a packet-group $\{P_{(4,5)}^{(2)}, P_{(2,3)}^{(3)}\}$ chosen from the caches of $\widetilde{\mathcal{M}}=\{2,3\}$ that each packet contains a fragment only requested by the user-groups in $\widetilde{\mathcal{M}}$, as seen from the second row in the cached packets. Then, in Step 2, select these $\widetilde{M}=|\widetilde{\mathcal{M}}|=2$ fragments overlined in the table and XOR them together to transmit $\overline{S_{5,(4,5)}^{(2)}}\oplus \overline{S_{3,(2,3)}^{(3)}}$, and meanwhile transmit the other $\widetilde{M}(\alpha-1)=2$ unselected fragments directly. Then, the delivery of the $\widetilde{M}\alpha=4$ fragments in the packet-group can be completed with a delivery load of 3, i.e., a delivery gain of 1. After that, there are two packets remaining in caches 2 and 3. Thus in Step 3, the server delivers $\alpha-1=1$ fragment for each packet by transmitting $S_{4,(1,4)}^{(2)}$ and $S_{2,(1,2)}^{(3)}$, which are requested by more user-groups than the remaining untransmitted ones. And the remaining ones will be delivered at the last stage. The total delivery load according to Type-\MyRoman{4} packets is 9. 
\end{example}
\par Next, we present the general delivery scheme according to Type-\MyRoman{4} packets.
\subsubsection{Step 1} Deliver the fragments encoded in each packet-group $\{P_{(n_{1}^{(m)},...,n_{\alpha}^{(m)})}^{(m)}:m\in\mathcal{M}_{\mathcal{V}}\}$ that corresponds to an $(\alpha+1)$-request set $\mathcal{V}$  satisfying (\ref{equ:PacketofType4_ConsiditionforStep1_a}). And the delivery scheme for the fragments encoded in  $\{P_{(n_{1}^{(m)},...,n_{\alpha}^{(m)})}^{(m)}:m\in\mathcal{M}_{\mathcal{V}}\}$ can be summarized by two sub-steps as follows:
\begin{itemize}
  \item Select $\alpha+1$ fragments from $\alpha+1$ packets from $\{P_{(n_{1}^{(m)},...,n_{\alpha}^{(m)})}^{(m)}:m\in\mathcal{M}_{\mathcal{V}}\}$ such that each fragment comes from a different file in $\mathcal{V}=\{r_{0},r_{1},...,r_{\alpha}\}$, which are denoted as
        \begin{equation}\label{equ:SelectedFragmentsforType4}
  \boxed{S_{r_{0},(n_{1}^{(m_{0})},...,n_{\alpha}^{(m_{0})})}^{(m_{0})}},~ \boxed{S_{r_{1},(n_{1}^{(m_{1})},...,n_{\alpha}^{(m_{1})})}^{(m_{1})}},...,~\boxed{S_{r_{\alpha}, (n_{1}^{(m_{\alpha})},...,n_{\alpha}^{(m_{\alpha})})}^{(m_{\alpha})}},
  \end{equation}
  where the box is adopted to differentiate with other $M_{\mathcal{V}}\alpha-(\alpha+1)$ unselected fragments in the packet-group. Note that the $\alpha+1$ fragments from certain $\alpha+1$ caches $m_{0},m_{1},...,m_{\alpha}\in\mathcal{M}_{\mathcal{V}}$ should always exist as we can have
$\bigcup\limits_{i=0}^{\alpha} \{n_{1}^{(m_{i})},...,n_{\alpha}^{(m_{i})}\}=\mathcal{V}$ and $(n_{1}^{(m_{0})},...,n_{\alpha}^{(m_{0})})\neq\cdots\neq(n_{1}^{(m_{\alpha})},...,n_{\alpha}^{(m_{\alpha})})$ according to (\ref{equ:PacketofType4_ConsiditionforStep1_a}). This can be proved by assuming $m_{i}\in\mathcal{M}(n_{m_{i}})$ and $m_{j}\in\mathcal{M}(n_{m_{j}})$, where $n_{m_{i}}\in\mathcal{D}_{m_{i}}\cap\mathcal{V}$ and $n_{m_{j}}\in\mathcal{D}_{m_{j}}\cap\mathcal{V}$ for $n_{m_{i}}\neq n_{m_{j}}$. Based on the $\alpha+1$ selected fragments given in (\ref{equ:SelectedFragmentsforType4}), the server can then XOR each unselected fragment together with a selected fragment of the same file to form $M_{\mathcal{V}}\alpha-(\alpha+1)$ pairwise-coded packets and then transmit them, which are
    \begin{equation}\label{equ:DeliveredPacketsType4_Step1}
    W_{r_{i}}^{(m)}=S_{r_{i},(n_{1}^{(m)},...,n_{\alpha}^{(m)})}^{(m)}\oplus \boxed{S_{r_{i},(n_{1}^{(m_{i})},...,n_{\alpha}^{(m_{i})})}^{(m_{i})}},~r_{i}\in\mathcal{V},
  \end{equation}
  where $m\in\mathcal{M}_{\mathcal{V}}\setminus\{m_{i}\}~\mathrm{such}~\mathrm{that} ~r_{i}\in\{n_{1}^{(m)},...,n_{\alpha}^{(m)}\}$ with $i=0,1,...,\alpha$.
  \item Transmit one packet encoded by the $\alpha+1$ selected fragments given in (\ref{equ:SelectedFragmentsforType4}), which is
     \begin{equation}\label{equ:DeliveredPacketsType4_Step2}
   W_{\mathcal{V}}=\bigoplus\limits_{r_{i}\in\mathcal{V}} \boxed{S_{r_{i},(n_{1}^{(m_{i})},...,n_{\alpha}^{(m_{i})})}^{(m_{i})}}=\bigoplus\limits_{i=0}^{\alpha} \boxed{S_{r_{i},(n_{1}^{(m_{i})},...,n_{\alpha}^{(m_{i})})}^{(m_{i})}}.
  \end{equation}
\end{itemize}
Based on the transmissions given in (\ref{equ:DeliveredPacketsType4_Step1}) and (\ref{equ:DeliveredPacketsType4_Step2}), all user-groups in $\mathcal{M}_{\mathcal{V}}$ can obtain their requested fragments encoded in $\{P_{(n_{1}^{(m)},...,n_{\alpha}^{(m)})}^{(m)}:m\in\mathcal{M}_{\mathcal{V}}\}$ via XOR operations. Take user-group $m\in\mathcal{M}_{\mathcal{V}}$ caching $P_{(n_{1}^{(m)},...,n_{\alpha}^{(m)})}^{(m)}$ for example. Assuming it requests $S_{r_{0}}$ with $\mathcal{D}_{m}\cap \mathcal{V}=\{r_{0}\}$, we have $\{n_{1}^{(m)},...,n_{\alpha}^{(m)}\}=\mathcal{V}\setminus\{r_{0}\}=\{r_{1},...,r_{\alpha}\}$ and $m\neq m_{0}$ as user-group $m_{0}$ caches $P_{(n_{1}^{(m_{0})},...,n_{\alpha}^{(m_{0})})}^{(m_{0})}$ containing the fragment of $S_{r_{0}}$. Thus user-group $m$ can first obtain
   \begin{equation}\label{equ:DeliveredPacketsType4_Process_Step2a}
  \begin{split}
&P_{(n_{1}^{(m)},...,n_{\alpha}^{(m)})}^{(m)}\oplus\left(\bigoplus\limits_{r_{i}\in\mathcal{V}\setminus\{r_{0}\}} W_{r_{i}}^{(m)}\right)\oplus W_{\mathcal{V}}\\
&=\left(\bigoplus\limits_{i=1}^{\alpha} S_{r_{i},(n_{1}^{(m)},...,n_{\alpha}^{(m)})}^{(m)}\right) \oplus \left(\bigoplus\limits_{i=1}^{\alpha} \left(S_{r_{i},(n_{1}^{(m)},...,n_{\alpha}^{(m)})}^{(m)}\oplus \boxed{S_{r_{i},(n_{1}^{(m_{i})},...,n_{\alpha}^{(m_{i})})}^{(m_{i})}}\right)\right)\oplus W_{\mathcal{V}}\\
    &=\bigoplus\limits_{i=1}^{\alpha} \boxed{S_{r_{i},(n_{1}^{(m_{i})},...,n_{\alpha}^{(m_{i})})}^{(m_{i})}} \oplus  \bigoplus\limits_{i=0}^{\alpha} \boxed{S_{r_{i},(n_{1}^{(m_{i})},...,n_{\alpha}^{(m_{i})})}^{(m_{i})}}=\boxed{S_{r_{0},(n_{1}^{(m_{0})},...,n_{\alpha}^{(m_{0})})}^{(m_{0})}},
    \end{split}
  \end{equation}
  which is the selected fragment of $S_{r_{0}}$ cached by user-group $m_{0}$.
As each unselected fragment of $S_{r_{0}}$ has been encoded with $\boxed{S_{r_{0},(n_{1}^{(m_{0})},...,n_{\alpha}^{(m_{0})})}^{(m_{0})}}$ and transmitted via (\ref{equ:DeliveredPacketsType4_Step1}), user-group $m$ can then obtain all the unselected fragments of $S_{r_{0}}$ by combining (\ref{equ:DeliveredPacketsType4_Step1}) with (\ref{equ:DeliveredPacketsType4_Process_Step2a}). Thus the delivery of the $M_{\mathcal{V}}\alpha$ fragments in $\{P_{(n_{1}^{(m)},...,n_{\alpha}^{(m)})}^{(m)}:m\in\mathcal{M}_{\mathcal{V}}\}$ can be completed since $\mathcal{M}_{\mathcal{V}}=\bigcup\limits_{r\in\mathcal{V}}\mathcal{M}(r)$ and $\mathcal{D}_{m}\cap\mathcal{V}=\emptyset$ for any $m\in\mathcal{M}\setminus\mathcal{M}_{\mathcal{V}}$.
\par According to the transmissions given in (\ref{equ:DeliveredPacketsType4_Step1}) and (\ref{equ:DeliveredPacketsType4_Step2}), the delivery load for each packet-group $\{P_{(n_{1}^{(m)},...,n_{\alpha}^{(m)})}^{(m)}:m\in\mathcal{M}_{\mathcal{V}}\}$ is $M_{\mathcal{V}}\alpha-(\alpha+1)+1=(M_{\mathcal{V}}-1)\alpha$ and thus the delivery gain is $\alpha$ since there are $M_{\mathcal{V}}\alpha$ fragments. This means no matter what $\mathcal{M}_{\mathcal{V}}$ is, the delivery gain for the corresponding packet-group is $\alpha$. To find all such packet-groups such that different groups consist of totally different packets, all the valid $(\alpha+1)$-request sets need to be found first, and thus the following additional condition needs to be added with (\ref{equ:PacketofType4_ConsiditionforStep1_a}) for any two obtained ($\alpha+1$)-request sets $\mathcal{V}_{1}$ and $\mathcal{V}_{2}$,
 \begin{equation}\label{equ:PacketofType4_ConsiditionforStep1_b}
  \begin{cases}
   |\mathcal{V}_{1}\cap\mathcal{V}_{2}|<\alpha;~\mathrm{or}\\
   |\mathcal{V}_{1}\cap\mathcal{V}_{2}|=\alpha~\mathrm{and}~\triangle\mathcal{V}\not\subseteq \mathcal{D}_{m},~\forall m\in \mathcal{M},\\
  \end{cases}
  \end{equation}
where $\triangle\mathcal{V}\triangleq\left(\mathcal{V}_{1}\cup\mathcal{V}_{2}\right)\left\backslash\vphantom{\left(\mathcal{V}_{1}
\cap\mathcal{V}\right)}\right.\left(\mathcal{V}_{1}\cap\mathcal{V}_{2}\right)$ denotes the set of different requests between $\mathcal{V}_{1}$ and $\mathcal{V}_{2}$. The condition on $\triangle\mathcal{V}$ for $|\mathcal{V}_{1}\cap\mathcal{V}_{2}|=\alpha$ indicates that any two different requests in $\mathcal{V}_{1}$ and $\mathcal{V}_{2}$ cannot simultaneously come from any one of the user-groups in $\mathcal{M}$. Otherwise, only one request set can be selected. The search procedure is summarized in \emph{Algorithm} \ref{alg:SerachingRequestSets} and is illustrated in \emph{Appendix} \ref{app:SearchingAlgorithm}. Then according to each obtained $\mathcal{V}$, both the requesting user-groups $\mathcal{M}_{\mathcal{V}}$ and the corresponding packet-group $\{P_{(n_{1}^{(m)},...,n_{\alpha}^{(m)})}^{(m)}:m\in\mathcal{M}_{\mathcal{V}}\}$ can be obtained based on (\ref{equ:PacketofType4_ConsiditionforStep1_a}). Summing up the delivery gains for all the valid request sets $\mathcal{V}$, the total delivery load for Step 1 can be finally obtained.
\begin{algorithm}[ht!]
\caption{Search procedure for the valid $(\alpha+1)$-request sets satisfying (\ref{equ:PacketofType4_ConsiditionforStep1_a})}\label{alg:SerachingRequestSets}
\hspace*{0.02in} {\bf Input:} $\mathcal{M}$, $\mathcal{N}$, $\mathcal{N}_{\mathrm{R}}$, $\{\mathcal{D}_{m},D_{m}:m\in\mathcal{M}\}$, $\alpha$ \\
\hspace*{0.02in} {\bf Output:} All the valid $(\alpha+1)$-request sets denoted by $\mathcal{R}$
\begin{algorithmic}[1]
\State \textbf{Initialization:} Set $\mathcal{R}=\emptyset$ and define $\mathcal{Z}_{\alpha}\triangleq\{m_{1},...,m_{\alpha}\}$ as a set of any $\alpha$ user-groups.
\For{$m_{0}=1:M-\alpha$}
\For{$\mathcal{Z}_{\alpha}\in \mathcal{M}\setminus\{1,2,...,m_{0}\}$}
\For{\{$n_{m_{i}}\in \mathcal{D}_{m_{i}}\}_{i=0}^{\alpha}$ such that (\ref{equ:PacketofType4_ConsiditionforStep1_a}) holds}
\State{$\mathcal{V}_{1}=\{n_{m_{0}},n_{m_{1}},...,n_{m_{\alpha}}\}$;}
\If{$\mathcal{V}_{1}$ and $\mathcal{V}_{2}$ satisfy (\ref{equ:PacketofType4_ConsiditionforStep1_b}),~$\forall~\mathcal{V}_{2}\in \mathcal{R}$}
 \State Update $\mathcal{R}=\{\mathcal{R},\mathcal{V}_{1}\}$;
 \EndIf
 \EndFor
\EndFor
\EndFor
\State \Return $\mathcal{R}$
\end{algorithmic}
\end{algorithm}
\subsubsection{Step 2} Deliver the fragments encoded in each packet-group $\{P_{(n_{1}^{(m)},...,n_{\alpha}^{(m)})}^{(m)}:m\in\widetilde{\mathcal{M}}\}$ for any $\widetilde{\mathcal{M}}\triangleq\{m_{0},m_{1},...,m_{\widetilde{M}-1}\}\subseteq\mathcal{M}$ over the remaining Type-\MyRoman{4} packets that satisfies
       \begin{equation}\label{equ:PacketsofType4_M2Condition}
  \{n_{1}^{(m)},...,n_{\alpha}^{(m)}\}\not\subseteq\bigcup\limits_{k\in\mathcal{M}\setminus\widetilde{\mathcal{M}}}
   \mathcal{D}_{k},~ m\in\widetilde{\mathcal{M}},
  \end{equation}
which indicates each packet should contain at least one fragment not requested by the user-groups in $\mathcal{M}\setminus\widetilde{\mathcal{M}}$. Note that $\widetilde{M}\geq 2$ and $(\ref{equ:PacketsofType4_M2Condition})$ can always be satisfied when $\widetilde{\mathcal{M}}=\mathcal{M}$. Then the delivery scheme for the fragments encoded in  $\{P_{(n_{1}^{(m)},...,n_{\alpha}^{(m)})}^{(m)}:m\in\widetilde{\mathcal{M}}\}$ can be summarized by two sub-steps as follows:
\begin{itemize}
  \item Transmit $\alpha-1$ fragments encoded in each packet $P_{(n_{1}^{(m_{i})},...,n_{\alpha}^{(m_{i})})}^{(m_{i})}$ except the fragment 
        \begin{equation}\label{equ:DeliveredCodedPacketsofType4_case2_Sub1}
 S_{n_{i},(n_{1}^{(m_{i})},...,n_{\alpha}^{(m_{i})})}^{(m_{i})},~n_{i}\not\in\bigcup\limits_{k\in\mathcal{M}\setminus\widetilde{\mathcal{M}} }\mathcal{D}_{k}~\mathrm{and}~n_{i}\in\{n_{1}^{(m_{i})},...,n_{\alpha}^{(m_{i})}\},
\end{equation}
where $i=0,1,...,\widetilde{M}-1$.
  \item XOR one untransmitted fragment $S_{n_{i},(n_{1}^{(m_{i})},...,n_{\alpha}^{(m_{i})})}^{(m_{i})}$ together with its next one from $i=0$ to $i=\widetilde{M}-2$ to form $\widetilde{M}-1$ pairwise-coded packets and transmit them, which are
\begin{equation}\label{equ:DeliveredCodedPacketsofType4_case2_Sub2}
 S_{n_{i},(n_{1}^{(m_{i})},...,n_{\alpha}^{(m_{i})})}^{(m_{i})}\oplus S_{n_{i+1},(n_{1}^{(m_{i+1})},...,n_{\alpha}^{(m_{i+1})})}^{(m_{i+1})},~i=0,1,...,\widetilde{M}-2.
\end{equation}
\end{itemize}
The directly transmitted $\alpha-1$ fragments in each packet can be delivered to the requesting user-groups inside both $\widetilde{\mathcal{M}}$ and $\mathcal{M}\setminus\widetilde{\mathcal{M}}$. Then we only need to consider the indirectly transmitted one fragment encoded in each packet. Assume user-group $m_{i}$ requests fragment $S_{n_{j},(n_{1}^{(m_{j})},...,n_{\alpha}^{(m_{j})})}^{(m_{j})}$, which is encoded in packet $P_{(n_{1}^{(m_{j})},...,n_{\alpha}^{(m_{j})})}^{(m_{j})}$ cached by user-group $m_{j}$ with $j>i$. Then user-group $m_{i}$ can first compute
\begin{equation}\label{equ:DeliveredCodedPacketsofType4_case2_Compute1}
\begin{split}
  &\sum_{l=i}^{j-1}\left(S_{n_{l},(n_{1}^{(m_{l})},...,n_{\alpha}^{(m_{l})})}^{(m_{l})}\oplus S_{n_{l+1},(n_{1}^{(m_{l+1})},...,n_{\alpha}^{(m_{l+1})})}^{(m_{l+1})}\right)
  =S_{n_{i},(n_{i}^{(m_{i})},...,n_{\alpha}^{(m_{i})})}^{(m_{i})}\oplus S_{n_{j},(n_{1}^{(m_{j})},...,n_{\alpha}^{(m_{j})})}^{(m_{j})}.
\end{split}
\end{equation}
After that user-group $m_{i}$ can obtain $S_{n_{j},(n_{1}^{(m_{j})},...,n_{\alpha}^{(m_{j})})}^{(m_{j})}$ via an XOR operation as it caches $S_{n_{i},(n_{i}^{(m_{i})},...,n_{\alpha}^{(m_{i})})}^{(m_{i})}$. In the same way user-group $m_{j}$ can also obtain $S_{n_{i},(n_{i}^{(m_{i})},...,n_{\alpha}^{(m_{i})})}^{(m_{i})}$ if it needs the fragment. Since $m_{i},m_{j}\in\widetilde{\mathcal{M}}$ are arbitrary, the selected fragments in each packet can be delivered to the requesting user-groups. Thus all the fragments in the packet-group can be delivered.
\par From the above, it can be seen that the delivery load for each packet-group satisfying (\ref{equ:PacketsofType4_M2Condition}) is $\widetilde{M}(\alpha-1)+\widetilde{M}-1=\widetilde{M}\alpha-1$. Thus no matter what $\widetilde{\mathcal{M}}$ is, the delivery gain for each packet-group is 1. And when $\alpha=1$, Steps 1 and 2 can be merged into one step as the delivery methods in both steps become the same. To find all the packet-groups, for each $\widetilde{M}=2,3,...,M$, set a reference cache $m_{0}$ by letting $m_{0}= 1,...,M-\widetilde{M}+1$, then given each reference cache $m_{0}$, find the remaining cached packets from cache $m_{0}$ and every other $\widetilde{M}-1$ caches from $\{m_{0}+1,...,M\}$ that can form packet-groups to satisfy (\ref{equ:PacketsofType4_M2Condition}). Note that (\ref{equ:PacketsofType4_M2Condition}) can always be guaranteed for any packet-group for $\widetilde{M}=M$. The search procedure for the packet-groups is summarized in \emph{Algorithm} \ref{alg:UsingM2forStep2_TypeIV}. Summing up the delivery gains for all packet-groups, the total delivery load for Step 2 can be finally obtained.
  \begin{algorithm}[ht!]
\caption{Search procedure for all the packet-groups satisfying (\ref{equ:PacketsofType4_M2Condition})}\label{alg:UsingM2forStep2_TypeIV}
\hspace*{0.02in} {\bf Input:} $\mathcal{M}$, $\mathcal{N}$, $\mathcal{N}_{\mathrm{R}}$, $\{\mathcal{D}_{m},D_{m}:m\in\mathcal{M}\}$, $\alpha$ \\
\hspace*{0.02in} {\bf Output:} All the valid packet-groups denoted by $\mathcal{G}$
\begin{algorithmic}[1]
\State \textbf{Initialization:} Set $\mathcal{G}=\emptyset$, and define $\mathcal{P}_{m}$ as the set of the remaining Type-\MyRoman{4} packets in cache $m\in\mathcal{M}$ after Step 1 and $\mathcal{Z}_{\widetilde{M}-1}\triangleq\{m_{1},...,m_{\widetilde{M}-1}\}$ as a set of any $\widetilde{M}-1$ caches.
\For{$\widetilde{M}=2:M$}
\For{$m_{0}=1:M-\widetilde{M}+1$}
\For{$\mathcal{Z}_{\widetilde{M}-1}\in\{m_{0}+1,...,M\}$}
\For{$\{P_{(n_{1}^{(m_{i})},...,n_{\alpha}^{(m_{i})})}^{(m_{i})}\in\mathcal{P}_{m_{i}}:m_{i}\in\widetilde{\mathcal{M}}=\{m_{0},\mathcal{Z}_{\widetilde{M}-1}\}\}$ such that (\ref{equ:PacketsofType4_M2Condition}) holds}
\State{$\mathcal{P}=\{P_{(n_{1}^{(m_{i})},...,n_{\alpha}^{(m_{i})})}^{(m_{i})}\in \mathcal{P}_{m_{i}}:m_{i}\in\widetilde{\mathcal{M}}\}$;}
\State Update $\mathcal{G}=\{\mathcal{G},\mathcal{P}\}$;
\State{Update $\{\mathcal{P}_{m}:m\in\mathcal{M}\}$ by $\mathcal{P}_{m_{i}}=\mathcal{P}_{m_{i}}\setminus\{P_{(n_{1}^{(m_{i})},...,n_{\alpha}^{(m_{i})})}^{(m_{i})}\}$,~$m_{i}\in\widetilde{\mathcal{M}}$;}
 \EndFor
\EndFor
\EndFor
\EndFor
\State \Return $\mathcal{G}$
\end{algorithmic}
\end{algorithm}
\subsubsection{Step 3} Decode the remaining Type-\MyRoman{4} packets in all the caches by transmitting $\alpha-1$ most requested fragments for each one. The untransmitted one fragment in each packet will be delivered at the last stage.
\par As the delivery gains for Steps 1 and 2 depend on the specific requests of the entire caching system, we use $\delta$ to denote the sum of them and use $N_{\mathrm{\MyRoman{4}}}^{(\mathrm{DEL})}$ to denote the corresponding number of Type-\MyRoman{4} packets delivered in these two steps. Then the number of transmissions in Step 3 is $\left(\sum\limits_{m=1}^{M}\textit{\textsf{C}}_{N_{\mathrm{R}}-D_{m}}^{\alpha}-N_{\mathrm{\MyRoman{4}}}^{(\mathrm{DEL})}\right)(\alpha-1)$ since the number of Type-\MyRoman{4} packets cached by user-group $m\in \mathcal{M}$ is $\textit{\textsf{C}}_{N_{\mathrm{R}}-D_{m}}^{\alpha}$. Thus the delivery load according to Type-\MyRoman{4} packets is given by
\begin{equation}\label{equ:NumofBroadcastforType4_Draft_sub1}
\begin{split}
  T_{\mathrm{\MyRoman{4}}} &=N_{\mathrm{\MyRoman{4}}}^{(\mathrm{DEL})}\alpha-\delta+\left(\sum\limits_{m=1}^{M}\textit{\textsf{C}}_{N_{\mathrm{R}}-D_{m}}^{\alpha}-N_{\mathrm{\MyRoman{4}}}^{(\mathrm{DEL})}\right)(\alpha-1)\\
  &=\sum\limits_{m=1}^{M}\textit{\textsf{C}}_{N_{\mathrm{R}}-D_{m}}^{\alpha}(\alpha-1)+N_{\mathrm{\MyRoman{4}}}^{(\mathrm{DEL})}-\delta.
\end{split}
\end{equation}
Since the total number of fragments in Type-\MyRoman{4} packets is $\sum\limits_{m=1}^{M}\textit{\textsf{C}}_{N_{\mathrm{R}}-D_{m}}^{\alpha} \alpha$, the number of remaining untransmitted fragments is
\begin{equation}\label{equ:RemaininguntransmittedFragments_Type4}
\begin{split}
  T_{\mathrm{\MyRoman{4}}}^{(\mathrm{RM})} &=\sum\limits_{m=1}^{M}\textit{\textsf{C}}_{N_{\mathrm{R}}-D_{m}}^{\alpha}-N_{\mathrm{\MyRoman{4}}}^{(\mathrm{DEL})}.
\end{split}
\end{equation}
\subsection{Last Stage}\label{sec:subsec:DeliveryLsatStage}
The last stage is to deliver all the $(T_{\mathrm{\MyRoman{2}}}^{(\mathrm{RM})}+T_{\mathrm{\MyRoman{3}}}^{(\mathrm{RM})}+T_{\mathrm{\MyRoman{4}}}^{(\mathrm{RM})})$ untransmitted fragments remaining in Types \MyRoman{2}, \MyRoman{3} and \MyRoman{4} packets that are requested by multiple user-groups, where $T_{\mathrm{\MyRoman{2}}}^{(\mathrm{RM})}$, $T_{\mathrm{\MyRoman{3}}}^{(\mathrm{RM})}$ and $T_{\mathrm{\MyRoman{4}}}^{(\mathrm{RM})}$ are given in (\ref{equ:RemaininguntransmittedFragments_Type2}),
(\ref{equ:RemaininguntransmittedFragments_Type3}) and (\ref{equ:RemaininguntransmittedFragments_Type4}), respectively. We first use the following example to illustrate the delivery scheme at the last stage.
\begin{table*} [ht]
  \centering
  \renewcommand\arraystretch{1.3}
\begin{tabular}{|c|c|c|c|c|}
\hline
  \multicolumn{2}{|c|}{User-group $m$} &$1$&$2$&$3$  \\
   \hline
   \multicolumn{2}{|c|}{Requested files} &$S_{1}$, $S_{2}$, $S_{3}$&$S_{2}$, $S_{3}$&$S_{1}$\\
   \hline
 \multirow{6}{*}{All cached}&\multirow{2}{*}{Type-\MyRoman{2}}&\multirow{2}{*}{$\left\backslash\vphantom{\frac{a^{\frac{1}{2}}}{b^{\frac{1}{2}}}}\right.$}&$S_{1,(1,2)}^{(2)}\oplus S_{2,(1,2)}^{(2)}$&$S_{1,(1,2)}^{(3)}\oplus S_{2,(1,2)}^{(3)}$ \\
   \multirow{5}{*}{packets}& &&$S_{1,(1,3)}^{(2)}\oplus S_{3,(1,3)}^{(2)}$&$S_{1,(1,3)}^{(3)}\oplus S_{3,(1,3)}^{(3)}$ \\
   \cline{2-5}
   &\multirow{3}{*}{Type-\MyRoman{3}}&$S_{1,(1,2)}^{(1)}\oplus S_{2,(1,2)}^{(1)}$&$S_{2,(2,3)}^{(2)}\oplus S_{3,(2,3)}^{(2)}$&\multirow{3}{*}{$\left\backslash\vphantom{\frac{a^{\frac{1}{2}}}{b^{\frac{1}{2}}}}\right.$}  \\
    & &$S_{1,(1,3)}^{(1)}\oplus S_{3,(1,3)}^{(1)}$& & \\
    & &$S_{2,(2,3)}^{(1)}\oplus S_{3,(2,3)}^{(1)}$& & \\
    \cline{2-5}
   & Type-\MyRoman{4}&$\left\backslash\vphantom{\frac{a^{1 }}{b^{1}}}\right.$&$\left\backslash\vphantom{\frac{a^{2}}{b^{1}}}\right.$&$S_{2,(2,3)}^{(3)}\oplus S_{3,(2,3)}^{(3)}$  \\
   \hline
   \multirow{3}{*}{Transmissions according }&Type-\MyRoman{2}&$\left\backslash\vphantom{\frac{a^{1 }}{b^{1}}}\right.$&$S_{1,(1,2)}^{(2)}$, $S_{1,(1,3)}^{(2)}$&$S_{2,(1,2)}^{(3)}$, $S_{3,(1,3)}^{(3)}$ \\
   \cline{2-5}
   \multirow{2}{*}{to different packet types}& Type-\MyRoman{3} &$S_{2,(1,2)}^{(1)}$, $S_{3,(1,3)}^{(1)}$, $S_{3,(2,3)}^{(1)}$ & $S_{2,(2,3)}^{(2)}$& $\left\backslash\vphantom{\frac{a^{1}}{b^{1}}}\right.$ \\
    \cline{2-5}
    &Type-\MyRoman{4}&$\left\backslash\vphantom{\frac{a^{1 }}{b^{1}}}\right.$&$\left\backslash\vphantom{\frac{a^{1 }}{b^{1}}}\right.$&$S_{3,(2,3)}^{(3)}$ \\
   \hline
   \multicolumn{2}{|c|}{Remaining fragments}&$S_{1,(1,2)}^{(1)}$, $S_{1,(1,3)}^{(1)}$, $S_{2,(2,3)}^{(1)}$&$S_{2,(1,2)}^{(2)}$, $S_{3,(1,3)}^{(2)}$, $S_{3,(2,3)}^{(2)}$&$S_{1,(1,2)}^{(3)}$, $S_{1,(1,3)}^{(3)}$, $S_{2,(2,3)}^{(3)}$ \\
   \hline
  \multicolumn{2}{|c|}{\multirow{3}{*}{Transmissions at the}}&\multicolumn{3}{c|}{$S_{1,(1,2)}^{(1)}\oplus S_{2,(1,2)}^{(2)}$, $S_{1,(1,2)}^{(1)}\oplus S_{1,(1,2)}^{(3)}$} \\
   \multicolumn{2}{|c|}{\multirow{2}{*}{last stage}}&\multicolumn{3}{c|}{$S_{1,(1,3)}^{(1)}\oplus S_{3,(1,3)}^{(2)}$, $ S_{1,(1,3)}^{(1)}\oplus S_{1,(1,3)}^{(3)}$} \\
   \multicolumn{2}{|c|}{}&\multicolumn{3}{c|}{$S_{2,(2,3)}^{(1)}\oplus S_{3,(2,3)}^{(2)}$, $S_{2,(2,3)}^{(1)}\oplus S_{2,(2,3)}^{(3)}$} \\
   \hline
\end{tabular}\caption{Illustration of the delivery scheme at the last stage.}\label{tab:DeliverySchemeofPacketType5}
\end{table*}
\begin{example}\label{exm:joint_delivery_phase}
 Assume $(N,M,D,\alpha)=(3,3,6,2)$, where $N_{\mathrm{R}}=N=3$, $\mathcal{D}_{1}=\{1,2,3\}$, $\mathcal{D}_{2}=\{2,3\}$ and $\mathcal{D}_{3}=\{1\}$. Then all the cached packets can be divided into three types as shown in Table \ref{tab:DeliverySchemeofPacketType5}. For Type-\MyRoman{2} packets, as user-group 1 requests all the files but its cache does not contain such type of packets, the server just transmits $\alpha-1=1$ fragment except the only one local fragment in each packet; for the similar reason for Type-\MyRoman{3} packets, the server just transmits $\alpha-1=1$ fragments except the least requested local fragment in each packet, where the least requested local fragments can be arbitrarily selected since the requests for each file are the same; for Type-\MyRoman{4} packets, as there is no packet-groups that can be delivered by Step 1 and Step 2, the server also just transmits $\alpha-1=1$ fragment for the only one packet in cache 3, where the transmitted fragment can be also arbitrarily selected. Thus, after the separate delivery of each type of packets, every cache still contains three untransmitted fragments. Then at the last stage, take the fragments in cache 1 as the reference fragments and XOR them together with the fragments in cache 2 and cache 3, respectively, to form 3 groups of packets, each group consisting of two pairwise-coded packets is transmitted to deliver the 3 fragments in them. And the final delivery gain is 3.
\end{example}
 \par Next, we present the general delivery scheme for the last stage, which is implemented by the following two steps:
\subsubsection{Step 1} Deliver the untransmitted fragments that are requested by multiple user-groups and can be pairwise-encoded among the $M$ caches similar to Step 2 in Type-\MyRoman{3} packets. Take the fragments from the cache having the minimum number of untransmitted fragments as the reference fragments. Then for each reference fragment, XOR it together with an unrepeatedly selected fragment from any other caches to form $M-1$ pairwise-code packets and transmit them. Then the fragments in these transmitted packets can be delivered to the requesting user-groups.
\subsubsection{Step 2} Deliver all the remaining fragments that are requested by multiple user-groups via direct transmissions.
\par Define $\Delta$ as the minimum number of the remaining untransmitted fragments over the $M$ caches, then the number of the transmitted pairwise-coded packets in Step 1 is $(M-1)\Delta$. Since each cache has been delivered $\Delta$ fragments after Step 1, the number of the transmitted fragments in Step 2 is $(T_{\mathrm{\MyRoman{2}}}^{(\mathrm{RM})}+T_{\mathrm{\MyRoman{3}}}^{(\mathrm{RM})}+T_{\mathrm{\MyRoman{4}}}^{(\mathrm{RM})}-M\Delta)$. Thus the total delivery load at the last stage is given by
\begin{equation}\label{equ:RemaininguntransmittedFragments_All}
\begin{split}
  T_{\mathrm{RM}}&=T_{\mathrm{\MyRoman{2}}}^{(\mathrm{RM})}+T_{\mathrm{\MyRoman{3}}}^{(\mathrm{RM})}+T_{\mathrm{\MyRoman{4}}}^{(\mathrm{RM})}-M\Delta+(M-1)\Delta\\
  &=T_{\mathrm{\MyRoman{2}}}^{(\mathrm{RM})}+T_{\mathrm{\MyRoman{3}}}^{(\mathrm{RM})}+T_{\mathrm{\MyRoman{4}}}^{(\mathrm{RM})}-\Delta,
\end{split}
\end{equation}
from which it can be seen that $\Delta$ also denotes the delivery gain obtained at the last stage.
\par Based on the delivery strategies introduced in Sections \ref{sec:subsec:DeliveryschemeforType1}-\ref{sec:subsec:DeliveryLsatStage}, we can summarize the proposed delivery scheme by Table \ref{tab:SummaryofDeliveryScheme}.
\begin{table*} [ht]
\centering
\renewcommand\arraystretch{1.3}
\begin{tabular}{|c|c|p{12.2cm}|}
\hline
\multicolumn{2}{|c|}{Delivery stages}& \multicolumn{1}{c|}{Specific delivery strategies} \\%
\hline
\multicolumn{2}{|c|}{Type-\MyRoman{1} cached packets}& Directly transmit the requested fragments encoded in each Type-\MyRoman{1} packet.  \\
\hline
\multicolumn{2}{|c|}{\multirow{3}{*}{Type-\MyRoman{2} cached packets}}& Step 1: Transmit $\alpha-1$ fragments encoded in each Type-\MyRoman{2} packet except the local fragment, which is assumed to be the fragment of $S_{n_{i}}$, $n_{i}\in \mathcal{N}_{\mathrm{R}}$;   \\
\cline{3-3}
\multicolumn{2}{|c|}{} & Step 2: Deliver the untransmitted local fragments of $S_{n_{i}}$ that can be pairwise-encoded among the caches of $\mathcal{M}(n_{i})$ , where $\mathcal{M}(n_{i})$ denotes the set of user-groups requesting $S_{n_{i}}$.   \\
\hline
\multicolumn{2}{|c|}{\multirow{3}{*}{Type-\MyRoman{3} cached packets}}& Step 1: Transmit $\alpha-1$ fragments encoded in each Type-\MyRoman{3} packet except the local fragment that is least requested by other user-groups;   \\
\cline{3-3}
\multicolumn{2}{|c|}{} & Step 2: Deliver the untransmitted local fragments that are requested by multiple user-groups and can be pairwise-encoded among the caches of $\mathcal{M}$.   \\
\hline
\multicolumn{2}{|c|}{\multirow{5}{*}{Type-\MyRoman{4} cached packets}}&Step 1: Deliver the fragments encoded in each packet-group that corresponds to an $(\alpha+1)$-request set $\mathcal{V}$  satisfying (\ref{equ:PacketofType4_ConsiditionforStep1_a}), where $2\leq\alpha\leq N$ and $\mathcal{V}$ is obtained by \emph{Algorithm} \ref{alg:SerachingRequestSets};   \\
\cline{3-3}
\multicolumn{2}{|c|}{}&Step 2: Deliver the fragments encoded in each packet-group over the remaining Type-\MyRoman{4} packets for any $\widetilde{\mathcal{M}}\subseteq\mathcal{M}$ that satisfies \ref{equ:PacketsofType4_M2Condition}, where the packet-group is obtained by \emph{Algorithm} \ref{alg:UsingM2forStep2_TypeIV};\\
\cline{3-3}
\multicolumn{2}{|c|}{}&Step 3: Decode the remaining Type-\MyRoman{4} packets in all the caches by transmitting $\alpha-1$ most requested fragments for each one.\\
\hline
\multicolumn{2}{|c|}{\multirow{4}{*}{Last stage}}& Step 1: Deliver the untransmitted fragments that are requested by multiple user-groups and can be pairwise-encoded among the $M$ caches similar to Step 2 in delivery of Type-\MyRoman{3} packets;   \\
\cline{3-3}
\multicolumn{2}{|c|}{} & Step 2: Deliver all the remaining fragments that are requested by multiple user-groups via direct transmissions.   \\
\hline
\end{tabular}\caption{Summary of the proposed delivery scheme.}\label{tab:SummaryofDeliveryScheme}
\end{table*}
\section{Analyses}\label{sec:Analyses}
\par In this section, we first show the correctness of our proposed delivery scheme and then summarize the overall delivery rate $R$. We also analyze the worst delivery rate $R^{*}\triangleq\max\limits_{\{\mathcal{D}_{m}:m\in\mathcal{M}\}} R$ and make a comparison with an existing uncoded prefetching scheme.
\subsection{Correctness}
\par Since all fragments of each file have been un-repeatedly encoded in the four types of cached packets, every file can be recovered by its requesting user-groups when all these encoded fragments are decoded and delivered to them. According to the proposed delivery scheme, the delivery of $S_{n}$ for any $n\in\mathcal{N}_{\mathrm{R}}$ can be summarized as follows:
\begin{itemize}
  \item In Type-\MyRoman{1} packets, all fragments of $S_{n}$ are directly transmitted one by one, thus all of them can be delivered to the user-groups in $\mathcal{M}(n)$;
  \item In Type-\MyRoman{2} packets, all fragments of $S_{n}$ in the caches of $\mathcal{M}\setminus\mathcal{M}(n)$ are not local fragments and each one belongs to the $\alpha-1$ directly transmitted fragments of a packet, thus they can be delivered to the user-groups in $\mathcal{M}(n)$ via direct transmissions; whereas all fragments of $S_{n}$ from the caches of $\mathcal{M}(n)$ for $|\mathcal{M}(n)|>1$ are local fragments and multi-requested, then the ones that can be pairwise-coded among $\mathcal{M}(n)$ are delivered to the user-groups in $\mathcal{M}(n)$ via coded transmissions and the remaining ones are left to be delivered at the last stage, where coded or direct transmissions or both may be used.
  \item  In Type-\MyRoman{3} packets, similar to Type-\MyRoman{2} packets, all fragments of $S_{n}$ in the caches of $\mathcal{M}\setminus\mathcal{M}(n)$ are not local fragments and can be delivered to the user-groups in $\mathcal{M}(n)$ via direct transmissions; whereas all fragments of $S_{n}$ in the caches of $\mathcal{M}(n)$ are local fragments, but different from Type-\MyRoman{2} packets, some of the local fragments of $S_{n}$ may have been delivered via direct $(\alpha-1)$-transmissions for each packet in Step 1 for Type-\MyRoman{3} packets, as every Type-\MyRoman{3} packet has more than one local fragment. Then part of the untransmitted local fragments of $S_{n}$ can be delivered by combining with the untransmitted local fragments of other files via pairwise-coded transmissions among the $M$ caches in Step 2 for Type-\MyRoman{3} packets. If there are still some fragments undelivered, then the remaining ones are left to be delivered at the last stage.
  \item In Type-\MyRoman{4} packets, the fragments of $S_{n}$ are only encoded in the caches of $\mathcal{M}\setminus\mathcal{M}(n)$ and can be delivered by Steps 1 and 2 if the corresponding condition is satisfied. Otherwise, they will be delivered via direct $(\alpha-1)$-transmissions for each packet in Step 3 for Type-\MyRoman{4} packets since they are not local fragments.
\end{itemize}
\par Hence all fragments of $S_{n}$ for any $n\in\mathcal{N}_{\mathrm{R}}$ can be delivered to the user-groups in $\mathcal{M}(n)$. Thus we can conclude that each user-group can obtain its requested files.
\subsection{Delivery Rate $R$}
Summing up $T_{\mathrm{\MyRoman{1}}}$, $T_{\mathrm{\MyRoman{2}}}$, $T_{\mathrm{\MyRoman{3}}}$, $T_{\mathrm{\MyRoman{4}}}$ and $T_{\mathrm{RM}}$ given in (\ref{equ:NumofBroadcastforType1_HMS}), (\ref{equ:NumofBroadcastforType2_HCMS}), (\ref{equ:NumofBroadcastforType2a_HCMS}), (\ref{equ:NumofBroadcastforType4_Draft_sub1}) and (\ref{equ:RemaininguntransmittedFragments_All}), respectively, and dividing them by $M\textit{\textsf{C}}_{N-1}^{\alpha-1}$, the delivery rate for the proposed approach is given by
\begin{equation}\label{equ:RateofDeliveryforEquCaches_HCMS}
\begin{split}
R=&\frac{T_{\mathrm{\MyRoman{1}}}+T_{\mathrm{\MyRoman{2}}}+T_{\mathrm{\MyRoman{2}}}^{(\mathrm{RM})}
+T_{\mathrm{\MyRoman{3}}}+T_{\mathrm{\MyRoman{3}}}^{(\mathrm{RM})}+T_{\mathrm{\MyRoman{4}}}
+T_{\mathrm{\MyRoman{4}}}^{(\mathrm{RM})}-\Delta}{M\textit{\textsf{C}}_{N-1}^{\alpha-1}}.
\end{split}
\end{equation}
\par Note that $T_{\mathrm{\MyRoman{1}}}=MN_{\mathrm{R}}\left(\textit{\textsf{C}}_{N-1}^{\alpha-1}-\textit{\textsf{C}}_{N_{\mathrm{R}}-1}^{\alpha-1}\right)$ and
\begin{equation}\label{equ:Somesubsumsofequations}
  \begin{array}{l}
~T_{\mathrm{\MyRoman{2}}}+T_{\mathrm{\MyRoman{2}}}^{(\mathrm{RM})}=\sum\limits_{m=1}^{M}\textit{\textsf{C}}_{N_{\mathrm{R}}-D_{m}}^{\alpha-1} D_{m}\alpha
-\sum\limits_{n\in \mathcal{N}_{\mathrm{R}}}\min\limits_{m\in\mathcal{M}(n)}\textit{\textsf{C}}_{N_{\mathrm{R}}-D_{m}}^{\alpha-1},\\
\begin{array}{rl}
T_{\mathrm{\MyRoman{3}}}+T_{\mathrm{\MyRoman{3}}}^{(\mathrm{RM})}= & \sum\limits_{m=1}^{M}\sum\limits_{n=2}^{\min\{D_{m},\alpha\}}\textit{\textsf{C}}_{D_{m}}^{n} \textit{\textsf{C}}_{N_{\mathrm{R}}-D_{m}}^{\alpha-n}(\alpha-1) \\
&+\sum\limits_{m=1}^{M}\sum\limits_{n=2}^{\min\{D_{m},\alpha\}}\textit{\textsf{C}}_{D_{m}-\sigma_{m}}^{n} \textit{\textsf{C}}_{N_{\mathrm{R}}-D_{m}}^{\alpha-n} \\
&-\min\limits_{m\in\mathcal{M}}\sum\limits_{n=2}^{\min\{D_{m},\alpha\}}\textit{\textsf{C}}_{D_{m}-\sigma_{m}}^{n} \textit{\textsf{C}}_{N_{\mathrm{R}}-D_{m}}^{\alpha-n},
\end{array}\\
~T_{\mathrm{\MyRoman{4}}}+T_{\mathrm{\MyRoman{4}}}^{(\mathrm{RM})}=\sum\limits_{m=1}^{M}\textit{\textsf{C}}_{N_{\mathrm{R}}-D_{m}}^{\alpha} \alpha-\delta,\\
~\sum\limits_{m=1}^{M}\sum\limits_{n=0}^{\min\{D_{m},\alpha\}}\textit{\textsf{C}}_{D_{m}}^{n}\textit{\textsf{C}}_{N_{\mathrm{R}}-D_{m}}^{\alpha-n} \alpha=\sum\limits_{m=1}^{M}\textit{\textsf{C}}_{N_{\mathrm{R}}}^{\alpha}\alpha=MN_{\mathrm{R}}\textit{\textsf{C}}_{N_{\mathrm{R}}-1}^{\alpha-1},
  \end{array}
\end{equation}
where the first equality in (\ref{equ:Somesubsumsofequations}) is obtained by $\sum\limits_{n\in \mathcal{N}_{\mathrm{R}}} \sum\limits_{m\in\mathcal{M}(n)}a_{m}=\sum\limits_{m=1}^{M}D_{m}a_{m}$. We have the following result on the delivery rate $R$.
\begin{theorem} \label{thm:theorem_CodedPrefetchingHCMS}
Consider an $(N,M,D,\alpha)$ caching network with $D=\sum\limits_{m=1}^{M} D_{m}$ and $C=\frac{N}{M\alpha}$, where $D_{m}$ denotes the number of distinct requests of user-group $m\in \mathcal{M}$ and $C$ denotes the size of each cache. Denoting the total set of distinct requests of the system as $\mathcal{N}_{\mathrm{R}}$ and $N_{\mathrm{R}}=|\mathcal{N}_{\mathrm{R}}|\leq N$, the delivery rate is given by
\begin{equation}\label{equ:RateofDeliveryforEquCaches_Theorem}
\begin{split}
R=&N_{\mathrm{R}}-\sum_{n\in\mathcal{N}_{\mathrm{R}}}\min\limits_{m\in\mathcal{M}(n)}\frac{\textit{\textsf{C}}_{N_{\mathrm{R}}-D_{m}}^{\alpha-1}}{M\textit{\textsf{C}}_{N-1}^{\alpha-1}}+\sum_{m=1}^{M}\sum\limits_{n=2}^{\min\{D_{m},\alpha\}}
  \frac{\left(\textit{\textsf{C}}_{ D_{m}-\sigma_{m}}^{n}-\textit{\textsf{C}}_{ D_{m}}^{n}\right)\textit{\textsf{C}}_{N_{\mathrm{R}}-D_{m}}^{\alpha-n}}{M\textit{\textsf{C}}_{N-1}^{\alpha-1}}
\\&-\min_{m\in\mathcal{M}}\sum\limits_{n=2}^{\min\{D_{m},\alpha\}}\frac{\textit{\textsf{C}}_{ D_{m}-\sigma_{m}}^{n}\textit{\textsf{C}}_{N_{\mathrm{R}}-D_{m}}^{\alpha-n}}{M\textit{\textsf{C}}_{N-1}^{\alpha-1}}-\frac{\delta+\Delta}{M\textit{\textsf{C}}_{N-1}^{\alpha-1}},
\end{split}
\end{equation}
where $\sigma_{m}$ denotes the number of the distinct files that are only requested by user-group $m$, $\mathcal{M}(n)$ denotes the set of the user-groups requesting $S_{n}$ for any $n\in\mathcal{N}_{\mathrm{R}}$, $\delta$ denotes the delivery gain according to Type-\MyRoman{4} packets and $\Delta$ denotes the delivery gain obtained at the last stage.
\end{theorem}
\subsection{Worst Delivery Rate $R^{*}$}\label{sec:DRAnalyses:subsec:WorstDeliveryRate}
\par Based on the caching model and the delivery scheme, it can be seen that the classification of the four types of the cached packets depends on the specific requests of the users of the whole caching network. When the request-file set $\mathcal{N}_{\mathrm{R}}$, i.e., $N_{\mathrm{R}}$, is given, all the cached packets can be non-overlappingly classified. However, when $N_{\mathrm{R}}$ increases, some Type-\MyRoman{1} packets will be included into other types of packets as the unrequested fragments encoded in them will be re-defined as requested fragments. Thus the delivery load will increase for the whole caching network since the transmissions for each type of packets are non-overlapping and the delivery load utilizing any packet of Type-\MyRoman{2}, \MyRoman{3} and \MyRoman{4} is not less than $\alpha-1$, whereas that utilizing any packet of
Type-\MyRoman{1} is not greater than $\alpha-1$. Then we can conclude that the worst delivery rate $R^{*}$ should be achieved at $N_{\mathrm{R}}=N$. In the following, we analyze $R^{*}$ at $N_{\mathrm{R}}=N$ according to the total number of the distinct requests $D$ for the whole caching network.
\subsubsection{The case of $D=M$}
As $\sum\limits_{m=1}^{M}D_{m}=D$, we have $D_{1}=D_{2}=\cdots=D_{M}=1$, which means each user-group has only one distinct request. Let $N_{\mathrm{R}}=N$. Then according to (\ref{equ:RateofDeliveryforEquCaches_Theorem}), $R^{*}$ can be formulated as
   \begin{equation}\label{equ:RateofDeliveryforEquCaches_Deliverytrade_Proof_Case1_Eq2}
\begin{split}
R^{*} =&N-\left(\sum_{n\in\mathcal{N}}
\frac{\textit{\textsf{C}}_{N-1}^{\alpha-1}}{M\textit{\textsf{C}}_{N-1}^{\alpha-1}}+\frac{\delta+\Delta}{M\textit{\textsf{C}}_{N-1}^{\alpha-1}}\right)=N-\frac{N \textit{\textsf{C}}_{N-1}^{\alpha-1}+\delta+\Delta}{M\textit{\textsf{C}}_{N-1}^{\alpha-1}},
      \end{split}
      \end{equation}
where $\delta$ denotes the delivery gain according to Type-\MyRoman{4} packets and $\Delta$ denotes the minimum number of the remaining untransmitted fragments over the $M$ caches at the last stage.
\par Since there are $N$ distinct requests and each user-group has only one distinct request, every $\alpha+1$ distinct requests can meet (\ref{equ:PacketofType4_ConsiditionforStep1_a}) and (\ref{equ:PacketofType4_ConsiditionforStep1_b}). Thus the number of the $(\alpha+1)$-request sets is
     \begin{equation}\label{equ:NumofAlphasets}
     \textit{\textsf{C}}_{N}^{\alpha+1}=\frac{N(N-\alpha)}{\alpha(\alpha+1)}\textit{\textsf{C}}_{N-1}^{\alpha-1}.
     \end{equation}
\par Note that for each Type-\MyRoman{4} packet $P_{(n_{1},...,n_{\alpha})}^{(m)}$, $\{n_{1},...,n_{\alpha}, \mathcal{D}_{m}\}$ is always an $(\alpha+1)$-request set $\mathcal{V}$ since $D_{m}=1$, thus it can be included into the packet-group corresponding to $\mathcal{V}$. Then all the fragments in Type-\MyRoman{4} packets can be delivered by Step 1, and $T_{\mathrm{\MyRoman{4}}}^{(\mathrm{RM})}=0$ with $T_{\mathrm{\MyRoman{4}}}^{(\mathrm{RM})}$ defined in (\ref{equ:RemaininguntransmittedFragments_Type4}). As each request set meeting (\ref{equ:PacketofType4_ConsiditionforStep1_a}) and (\ref{equ:PacketofType4_ConsiditionforStep1_b}) can achieve a delivery gain of $\alpha$, based on (\ref{equ:NumofAlphasets}) $\delta$ is given by
\begin{equation}\label{equ:WorstDeliveryRate_Case1_Delta1}
\begin{split}
  \delta=\frac{N(N-\alpha)}{\alpha+1}\textit{\textsf{C}}_{N-1}^{\alpha-1}.
  \end{split}
\end{equation}
As $D_{1}=D_{2}=\cdots=D_{M}=1$, we have $T_{\mathrm{\MyRoman{2}}}^{(\mathrm{RM})}=0$ and $T_{\mathrm{\MyRoman{3}}}^{(\mathrm{RM})}=0$, where $T_{\mathrm{\MyRoman{2}}}^{(\mathrm{RM})}$ and $T_{\mathrm{\MyRoman{3}}}^{(\mathrm{RM})}$ are given in (\ref{equ:RemaininguntransmittedFragments_Type2}) and
(\ref{equ:RemaininguntransmittedFragments_Type3}), respectively. Thus we have $\Delta=0$. Substituting (\ref{equ:WorstDeliveryRate_Case1_Delta1}) into (\ref{equ:RateofDeliveryforEquCaches_Deliverytrade_Proof_Case1_Eq2}) leads to $R^{*}=N-\frac{N(N+1)}{(\alpha+1)M}$. To further illustrate (\ref{equ:NumofAlphasets}) and (\ref{equ:WorstDeliveryRate_Case1_Delta1}), we present an example in Table \ref{tab:Type3ExampleforCase1}.
\begin{example}
Consider $(N,M,D,\alpha)=(4,5,5,2)$. For simplicity, we use the XORed file indices to denote the stored Type-\MyRoman{4} packets. For example, $(2\oplus3)^{(1)}$ denotes packet $S_{2,(2,3)}^{(1)}\oplus S_{3,(2,3)}^{(1)}$. Based on the requests given in the table, it can be seen that the $(\alpha+1)$-request sets that satisfy (\ref{equ:PacketofType4_ConsiditionforStep1_a}) and (\ref{equ:PacketofType4_ConsiditionforStep1_b}) are $\{1,2,3\}$, $\{1,2,4\}$, $\{1,3,4\}$ and $\{2,3,4\}$, respectively, which is equivalent to $\sum\limits_{n=1}^{N-\alpha}\textit{\textsf{C}}_{N-n}^{\alpha}=4$. As Table \ref{tab:DeliverySchemeofPacketType4} has already presented a detailed example on Step 1, we can easily verify that all Type-\MyRoman{4} packets can be decoded and delivered and $\delta=8$ according to the 4 $(\alpha+1)$-request sets shown in Table \ref{tab:Type3ExampleforCase1}, and the specific delivery process is omitted here.
\begin{table*} [ht]
  \centering
\begin{tabular}{|c|c|c|c|c|c|c|}
\hline
   User-group $m$&$1$&$2$&3&$4$&$5$ \\
   \hline
     Requested files&~~~~~~~~~~$S_{1}$~~~~~~~~~~&~~~~~~~~~~$S_{1}$~~~~~~~~~~&~~~~~~~~~~$S_{2}$~~~~~~~~~~&~~~~~~~~~~$S_{3}$~~~~~~~~~~&$S_{4}$ \\
    \hline
    \multirow{3}{*}{Cached Packets}& $(2\oplus 3)^{(1)}$ &$(2\oplus 3)^{(2)}$ & $(1\oplus 3)^{(3)}$&$(1\oplus 2)^{(4)}$&$(1\oplus 2)^{(5)}$ \\
    \multirow{2}{*}{(Type-\MyRoman{4})}& $(2\oplus 4)^{(1)}$ &$(2\oplus 4)^{(2)}$ & $(1\oplus 4)^{(3)}$&$(1\oplus 4)^{(4)}$&$(1\oplus 3)^{(5)}$ \\
    & $(3\oplus 4)^{(1)}$ &$(3\oplus 4)^{(2)}$ & $(3\oplus 4)^{(3)}$&$(2\oplus 4)^{(4)}$&$(2\oplus 3)^{(5)}$ \\
    \hline
     \multirow{4}{*}{Delivery using}&\multicolumn{5}{c|}{$\mathcal{V}=\{1,2,3\}$, $\mathcal{M}_{\mathcal{V}}=\{1,2,3,4\}$, packet-group:$\left\{(2\oplus 3)^{(1)},(2\oplus 3)^{(2)},(1\oplus 3)^{(3)},(1\oplus 2)^{(4)}\right\}$} \\
\cline{2-6}
      \multirow{3}{*}{Step 1}&\multicolumn{5}{c|}{$\mathcal{V}=\{1,2,4\}$, $\mathcal{M}_{\mathcal{V}}=\{1,2,3,5\}$, packet-group:$\left\{(2\oplus 4)^{(1)},(2\oplus 4)^{(2)},(1\oplus 4)^{(3)},(1\oplus 2)^{(5)}\right\}$} \\
      \cline{2-6}
       &\multicolumn{5}{c|}{$\mathcal{V}=\{1,3,4\}$, $\mathcal{M}_{\mathcal{V}}=\{1,2,4,5\}$, packet-group:$\left\{(3\oplus 4)^{(1)},(3\oplus 4)^{(2)},(1\oplus 4)^{(4)},(1\oplus 3)^{(5)}\right\}$} \\
       \cline{2-6}
      &\multicolumn{5}{l|}{$\mathcal{V}=\{2,3,4\}$, $\mathcal{M}_{\mathcal{V}}=\{3,4,5\}$, packet-group:$\left\{(3\oplus 4)^{(3)},(2\oplus 4)^{(4)},(2\oplus 3)^{(5)}\right\}$} \\
      \hline
\end{tabular}\caption{An example of delivery gain that meets (\ref{equ:NumofAlphasets}) and (\ref{equ:WorstDeliveryRate_Case1_Delta1}) for Type-\MyRoman{4} packets.}\label{tab:Type3ExampleforCase1}
\end{table*}
\end{example}
\par Based on the above analysis we then have the following corollary.
\begin{corollary}\label{cor:WorstDeliveryRate_Case1}
Consider an $(N,M,D,\alpha)$ caching network with $D=\sum\limits_{m=1}^{M} D_{m}$ and $C=\frac{N}{M\alpha}$, where $D_{m}$ denotes the number of distinct requests of user-group $m\in \mathcal{M}$ and $C$ denotes the size of each cache. For $D=M$, the worst delivery rate $R^{*}$ is given by
\begin{equation}\label{equ:WorstDeliveryRate_Case1}
  R^{*}=N-\frac{N(N+1)}{(\alpha+1)M},~M\geq N.
\end{equation}
\end{corollary}
\par Note that $M\geq N$ in \emph{Corollary} \ref{cor:WorstDeliveryRate_Case1} results from $D\geq N$ and the worst delivery rate for $D=M$ is the same as that given in \cite{Gomez-Vilardebo2016}. Thus our model and delivery approach can incorporate the existing results as a special case.
\subsubsection{The case of $D>M$} Then there is at least one user-group that has more than one request. Since the minimum number of Type-\MyRoman{4} packets over the $M$ caches is $\min\limits_{m\in\mathcal{M}}\textit{\textsf{C}}_{N-D_{m}}^{\alpha}$ under $N_{\mathrm{R}}=N$, there is at least a delivery gain of $\min\limits_{m\in\mathcal{M}}\textit{\textsf{C}}_{N-D_{m}}^{\alpha}$ that can be obtained using Step 2 in Type-\MyRoman{4} packets for $\widetilde{\mathcal{M}}=\mathcal{M}$. Assuming that no $\alpha+1$-request set $\mathcal{V}$ satisfies (\ref{equ:PacketofType4_ConsiditionforStep1_a}) for any $\mathcal{M}_{\mathcal{V}} \subseteq\mathcal{M}$ or packet-group satisfies (\ref{equ:PacketsofType4_M2Condition}) for any $\widetilde{\mathcal{M}}\subset\mathcal{M}$, the total delivery gain using Type-\MyRoman{4} packets can be formulated as
\begin{equation}\label{equ:cor:ProofofCorollary_Case2_1}
  \delta=\min\limits_{m\in\mathcal{M}}\textit{\textsf{C}}_{N-D_{m}}^{\alpha}.
\end{equation}
To verify (\ref{equ:cor:ProofofCorollary_Case2_1}) is achievable, we present an example in Table \ref{tab:Type3ExampleforCase2}.
 \begin{example}
Consider $(N,M,D,\alpha)=(6,4,12,3)$. Similarly, for simplicity, we use the XORed file indices to denote the stored Type-\MyRoman{4} packets. For example, $(3\oplus5\oplus6)^{(1)}$ denotes packet $S_{3,\{3,5,6\}}^{(1)}\oplus S_{5,\{3,5,6\}}^{(1)}\oplus S_{6,\{3,5,6\}}^{(1)}$. According to the requests and packets presented, there is no $\alpha+1$-request set $\mathcal{V}$ satisfies (\ref{equ:PacketofType4_ConsiditionforStep1_a}) for any $\mathcal{M}_{\mathcal{V}} \subseteq\mathcal{M}$ or packet-group satisfies (\ref{equ:PacketsofType4_M2Condition}) for any $\widetilde{\mathcal{M}}\subset\mathcal{M}$. Then the delivery gain utilizing Type-\MyRoman{4} packets mainly comes from Step 2, which is used to deliver the fragments in the $\min\limits_{m\in\mathcal{M}}\textit{\textsf{C}}_{N-D_{m}}^{\alpha}$ packet-groups over $\widetilde{\mathcal{M}}=\mathcal{M}$. Then (\ref{equ:cor:ProofofCorollary_Case2_1}) is verified.
\begin{table*} [ht]
  \centering
\begin{tabular}{|c|c|c|c|c|}
\hline
 User-group $m$&$1$&$2$&$3$&$4$ \\
 \hline
 Requested files &~~~~~~~$S_{1}$, $S_{2}$, $S_{4}$~~~~~~~&~~~~~~~$S_{3}$, $S_{5}$, $S_{6}$~~~~~~~&~~~~~~~$S_{1}$, $S_{2}$, $S_{4}$~~~~~~~&$S_{3}$, $S_{5}$, $S_{6}$\\
 \hline
 \multirow{2}{*}{Cached Packets}&\multirow{2}{*}{$(3\oplus 5\oplus6)^{(1)}$}&\multirow{2}{*}{$(1\oplus 2\oplus4)^{(2)}$}&\multirow{2}{*}{$(3\oplus 5\oplus6)^{(3)}$}&\multirow{2}{*}{$(1\oplus 2\oplus4)^{(4)}$}\\
\multirow{1}{*}{(Type-\MyRoman{4})}& & & & \\
    \hline
  Delivery using Step 2&\multicolumn{4}{|c|}{$\widetilde{\mathcal{M}}=\mathcal{M}$, $\min\limits_{m\in\mathcal{M}}\textit{\textsf{C}}_{N-D_{m}}^{\alpha}=1$ packet-group consisting of 4 packets from the 4 caches.}\\
    \hline
\end{tabular}\caption{An example of Type-\MyRoman{4} packets that meets (\ref{equ:cor:ProofofCorollary_Case2_1}). }\label{tab:Type3ExampleforCase2}
\end{table*}
\end{example}
\par Substituting (\ref{equ:cor:ProofofCorollary_Case2_1}) into (\ref{equ:RateofDeliveryforEquCaches_Theorem}) and letting $N_{\mathrm{R}}=N$, $R^{*}$ can be formulated as
    \begin{equation}\label{equ:cor:ProofofCorollary_Case2_2}
\begin{split}
R^{*}=&N-\sum_{n\in\mathcal{N}}\min\limits_{m\in\mathcal{M}(n)}\frac{\textit{\textsf{C}}_{N-D_{m}}^{\alpha-1}}{M\textit{\textsf{C}}_{N-1}^{\alpha-1}} +\sum_{m=1}^{M}\sum\limits_{n=2}^{\min\{D_{m},\alpha\}}\frac{\left(\textit{\textsf{C}}_{D_{m}-\sigma_{m}}^{n}-\textit{\textsf{C}}_{D_{m}}^{n}\right)
\textit{\textsf{C}}_{N-D_{m}}^{\alpha-n}}{M\textit{\textsf{C}}_{N-1}^{\alpha-1}}\\
&-\min_{m\in\mathcal{M}}\sum\limits_{n=2}^{\min\{D_{m},\alpha\}}\frac{\textit{\textsf{C}}_{D_{m}-\sigma_{m}}^{n}\textit{\textsf{C}}_{N-D_{m}}^{\alpha-n}}{M\textit{\textsf{C}}_{N-1}^{\alpha-1}}-\min\limits_{m\in\mathcal{M}} \frac{\textit{\textsf{C}}_{N-D_{m}}^{\alpha}}{M\textit{\textsf{C}}_{N-1}^{\alpha-1}}-\frac{\Delta}{M\textit{\textsf{C}}_{N-1}^{\alpha-1}},
      \end{split}
      \end{equation}
where $\mathcal{M}(n)$ denotes the set of the user-groups requesting $S_{n}$ for any $n\in\mathcal{N}$; $\sigma_{m}$ denotes the number of distinct files only requested by user-group $m$; and $\Delta$ denotes the minimum number of remaining fragments over the $M$ caches at the last stage.
\par Observing the terms related to $\{\sigma_{m}:m\in\mathcal{M}\}$ on the right-hand side of (\ref{equ:cor:ProofofCorollary_Case2_2}), it is seen that the largest $R^{*}$ is achieved at $\sigma_{m}=0$ for all $m\in\mathcal{M}$. This means that the worst delivery rate will be maximized at the point that every file is requested by multiple user-groups. Note that $\{\sigma_{m}=0:m\in\mathcal{M}\}$ does not depend on the value of $\delta$ in (\ref{equ:cor:ProofofCorollary_Case2_1}), $R^{*}$ can be further simplified as
\begin{equation}\label{equ:cor:ProofofCorollary_Case2_2a}
R^{*}=N-\sum_{n\in\mathcal{N}}\min\limits_{m\in\mathcal{M}(n)}\frac{\textit{\textsf{C}}_{N-D_{m}}^{\alpha-1}}{M\textit{\textsf{C}}_{N-1}^{\alpha-1}}-\min_{m\in\mathcal{M}}\sum\limits_{n=2}^{\min\{D_{m},\alpha\}}
\frac{\textit{\textsf{C}}_{D_{m}}^{n}\textit{\textsf{C}}_{N-D_{m}}^{\alpha-n}}{M\textit{\textsf{C}}_{N-1}^{\alpha-1}}-\min\limits_{m\in\mathcal{M}}
      \frac{\textit{\textsf{C}}_{N-D_{m}}^{\alpha}}{M\textit{\textsf{C}}_{N-1}^{\alpha-1}}-\frac{\Delta}{M\textit{\textsf{C}}_{N-1}^{\alpha-1}}.
      \end{equation}
\par Assume $D_{m_{1}}\geq D_{m_{2}}\geq\cdots\geq D_{m_{I+1}}\geq D_{m}$ for any $m\in\mathcal{M}\setminus\{m_{1},...,m_{I+1}\}$, where $I$ is the integer such that $N\geq \sum\limits_{i=1}^{I}D_{m_{i}}$ and $N<\sum\limits_{i=1}^{I+1}D_{m_{i}}$. Then, we have
\begin{equation}\label{equ:cor:ProofofCorollary_Case2_4}
\sum_{n\in\mathcal{N}} \min\limits_{m\in\mathcal{M}(n)}\frac{\textit{\textsf{C}}_{N-D_{m}}^{\alpha-1}}{M\textit{\textsf{C}}_{N-1}^{\alpha-1}}\geq \sum\limits_{i=1}^{I} \frac{\textit{\textsf{C}}_{N-D_{m_{i}}}^{\alpha-1}D_{m_{i}}}{M\textit{\textsf{C}}_{N-1}^{\alpha-1}}+\left(N-\sum\limits_{i=1}^{I}D_{m_{i}}\right)\frac{\textit{\textsf{C}}_{N-D_{m_{I+1}}}^{\alpha-1}}{M\textit{\textsf{C}}_{N-1}^{\alpha-1}},
\end{equation}
where the equality is achievable when $\bigcup\limits_{i=1}^{I}\mathcal{D}_{m_{i}}\subseteq \mathcal{N}$ and $\mathcal{N}\subset\bigcup\limits_{i=1}^{I+1}\mathcal{D}_{m_{i}}$, which indicates that the reference fragments for all $n\in\mathcal{N}$ in Step 2 for Type-\MyRoman{2} packets are chosen from caches $m_{1},...,m_{I+1}$ and the total number is $\sum\limits_{i=1}^{I} \textit{\textsf{C}}_{N-D_{m_{i}}}^{\alpha-1}D_{m_{i}}+\left(N-\sum\limits_{i=1}^{I}D_{m_{i}}\right)\textit{\textsf{C}}_{N-D_{m_{I+1}}}^{\alpha-1}$.
\par Combining (\ref{equ:cor:ProofofCorollary_Case2_2a}) with (\ref{equ:cor:ProofofCorollary_Case2_4}) we can further bound the worst delivery rate $R^{*}$ as
\begin{equation}\label{equ:cor:ProofofCorollary_Case2_WorstRate1}
\begin{split}
R^{*}\leq&N-\left(\sum\limits_{i=1}^{I} \frac{\textit{\textsf{C}}_{N-D_{m_{i}}}^{\alpha-1}D_{m_{i}}}{M\textit{\textsf{C}}_{N-1}^{\alpha-1}}+\left(N-\sum\limits_{i=1}^{I}D_{m_{i}}\right)\frac{\textit{\textsf{C}}_{N-D_{m_{I+1}}}^{\alpha-1}}{M\textit{\textsf{C}}_{N-1}^{\alpha-1}}\right.\\
&\left.+\min_{m\in\mathcal{M}}\sum\limits_{n=2}^{\min\{D_{m},\alpha\}}
\frac{\textit{\textsf{C}}_{D_{m}}^{n}\textit{\textsf{C}}_{N-D_{m}}^{\alpha-n}}{M\textit{\textsf{C}}_{N-1}^{\alpha-1}}+\min\limits_{m\in\mathcal{M}}
      \frac{\textit{\textsf{C}}_{N-D_{m}}^{\alpha}}{M\textit{\textsf{C}}_{N-1}^{\alpha-1}}+\frac{\Delta}{M\textit{\textsf{C}}_{N-1}^{\alpha-1}}\right)\\
      \triangleq& N-\left(G+\frac{\Delta}{M\textit{\textsf{C}}_{N-1}^{\alpha-1}}\right).
      \end{split}
      \end{equation}
\par Note that $D_{m}\textit{\textsf{C}}_{N-D_{m}}^{\alpha-1}$, $\sum\limits_{n=2}^{\min\{D_{m},\alpha\}}\textit{\textsf{C}}_{D_{m}}^{n}\textit{\textsf{C}}_{N-D_{m}}^{\alpha-n}$ and $\textit{\textsf{C}}_{N-D_{m}}^{\alpha}$ denote the numbers of Type-\MyRoman{2}, \MyRoman{3} and \MyRoman{4} packets in cache $m\in\mathcal{M}$, respectively, where the sum is $\textit{\textsf{C}}_{N}^{\alpha}$. According to the delivery scheme, it can be seen from (\ref{equ:cor:ProofofCorollary_Case2_WorstRate1}) that $\min\limits_{m\in\mathcal{M}}\sum\limits_{n=2}^{\min\{D_{m},\alpha\}}
 \textit{\textsf{C}}_{D_{m}}^{n}\textit{\textsf{C}}_{N-D_{m}}^{\alpha-n}$ and $\min\limits_{m\in\mathcal{M}}
      \textit{\textsf{C}}_{N-D_{m}}^{\alpha}$ are the numbers of delivered Types \MyRoman{3} and \MyRoman{4} packets in each cache, respectively, where the delivered packets represent the packets whose encoded fragments have been totally delivered; while $D_{m_{i}}\textit{\textsf{C}}_{N-D_{m_{i}}}^{\alpha-1}$, $i=1,...,I$ are the numbers of delivered Type-\MyRoman{2} packets in caches $m_{1},m_{2}...,m_{I}$, which indicates the Type-\MyRoman{2} packets in the $I$ caches that have been totally delivered. To obtain $\Delta$, we need to compute the minimum number of undelivered packets over the $M$ caches after the separate delivery of the four types of packets. Although the numbers of delivered Type-\MyRoman{2} packets in the other $M-I$ caches are not known, denoting $A_{m}$ as the number of delivered Type-\MyRoman{2} packets in cache $m\in\mathcal{M}$, we have $A_{m_{i}}=D_{m_{i}}\textit{\textsf{C}}_{N-D_{m_{i}}}^{\alpha-1}$ for $i\in\{1,2,...,I\}$ and $A_{m}\leq \sum\limits_{i=1}^{I}D_{m_{i}}\textit{\textsf{C}}_{N-D_{m_{i}}}^{\alpha-1}+\left(N-\sum\limits_{i=1}^{I}D_{m_{i}}\right) \textit{\textsf{C}}_{N-D_{m_{I+1}}}^{\alpha-1}$ for any $m\in\mathcal{M}\setminus\{m_{1},m_{2},...,m_{I}\}$ as the number of the delivered Type-\MyRoman{2} packets in cache $m$ should be not greater than the total number of reference fragments chosen from caches $m_{1},...,m_{I+1}$ in the delivery of Step 2 for Type-\MyRoman{2} packets. Then according to the definition of $\Delta$, we have
 \begin{equation}\label{equ:NumberofMinRemainingfragments_Case2}
\begin{split}
\Delta=&\min\limits_{j\in\mathcal{M}}\left(\textit{\textsf{C}}_{N}^{\alpha}-A_{j}-\min\limits_{m\in\mathcal{M}}\sum\limits_{n=2}^{\min\{D_{m},\alpha\}}
 \textit{\textsf{C}}_{D_{m}}^{n}\textit{\textsf{C}}_{N-D_{m}}^{\alpha-n}-\min\limits_{m\in\mathcal{M}}
      \textit{\textsf{C}}_{N-D_{m}}^{\alpha}\right)\\
=&\min\limits_{m\in\mathcal{M}}\left(\textit{\textsf{C}}_{N}^{\alpha}-A_{m}\right)-\min\limits_{m\in\mathcal{M}}\sum\limits_{n=2}^{\min\{D_{m},\alpha\}}
 \textit{\textsf{C}}_{D_{m}}^{n}\textit{\textsf{C}}_{N-D_{m}}^{\alpha-n}-\min\limits_{m\in\mathcal{M}}
      \textit{\textsf{C}}_{N-D_{m}}^{\alpha}\\
\geq&\left(\textit{\textsf{C}}_{N}^{\alpha}-\max\limits_{m\in \mathcal{M}}A_{m}-\min\limits_{m\in\mathcal{M}}\sum\limits_{n=2}^{\min\{D_{m},\alpha\}}
 \textit{\textsf{C}}_{D_{m}}^{n}\textit{\textsf{C}}_{N-D_{m}}^{\alpha-n}-\min\limits_{m\in\mathcal{M}}
      \textit{\textsf{C}}_{N-D_{m}}^{\alpha}\right)^{+}\\
\geq&\left(\textit{\textsf{C}}_{N}^{\alpha}-\sum\limits_{i=1}^{I}D_{m_{i}}\textit{\textsf{C}}_{N-D_{m_{i}}}^{\alpha-1}-\left(N-\sum\limits_{i=1}^{I}D_{m_{i}}\right) \textit{\textsf{C}}_{N-D_{m_{I+1}}}^{\alpha-1}\vphantom{-\min\limits_{m\in\mathcal{M}}\sum\limits_{n=2}^{\min\{D_{m},\alpha\}}
 \textit{\textsf{C}}_{D_{m}}^{n}\textit{\textsf{C}}_{N-D_{m}}^{\alpha-n}}\right.\\
&\left.-\min\limits_{m\in\mathcal{M}}\sum\limits_{n=2}^{\min\{D_{m},\alpha\}}
 \textit{\textsf{C}}_{D_{m}}^{n}\textit{\textsf{C}}_{N-D_{m}}^{\alpha-n}-\min\limits_{m\in\mathcal{M}}
      \textit{\textsf{C}}_{N-D_{m}}^{\alpha}\right)^{+}=\left(\textit{\textsf{C}}_{N}^{\alpha}-GM\textit{\textsf{C}}_{N-1}^{\alpha-1}\right)^{+},
      \end{split}
      \end{equation}
      where $G$ is defined in (\ref{equ:cor:ProofofCorollary_Case2_WorstRate1}). We can easily prove that the equality in (\ref{equ:NumberofMinRemainingfragments_Case2}) can be achieved by simply letting $D_{1}=N$ and $D_{2}=1$. Then $G=0$ and $\Delta=\textit{\textsf{C}}_{N}^{\alpha}$, and the delivery scheme is equivalent to first transmitting $\alpha-1$ fragments for each packet of Type-\MyRoman{2}, Type-\MyRoman{3} and Type-\MyRoman{4}, and then transmitting the remaining fragments by pairwise coding. This can be verified by \emph{Example} \ref{exm:joint_delivery_phase}. Based on (\ref{equ:cor:ProofofCorollary_Case2_WorstRate1}) and (\ref{equ:NumberofMinRemainingfragments_Case2}), we have the following corollary.
\begin{corollary}\label{cor:WorstDeliveryRate_Case2}
Consider an $(N,M,D,\alpha)$ caching network with $D=\sum\limits_{m=1}^{M} D_{m}$ and $C=\frac{N}{M\alpha}$, where $D_{m}$ denotes the number of distinct requests of user-group $m\in \mathcal{M}$ and $C$ denotes the size of each cache. For $D>M$, the worst delivery rate $R^{*}$ is given by
\begin{equation}\label{equ:cor:WorstDeliveryRate_Case2}
\begin{split}
 R^{*}=&N-\left(G+\left(\frac{N}{M\alpha}-G\right)^{+}\right),
      \end{split}
\end{equation}
where
\begin{equation}\label{equ:ParametersoftheEquation}
\begin{split}
   G=&\sum\limits_{i=1}^{I} \frac{\textit{\textsf{C}}_{N-D_{m_{i}}}^{\alpha-1}D_{m_{i}}}{M\textit{\textsf{C}}_{N-1}^{\alpha-1}}
   +\left(N-\sum\limits_{i=1}^{I}D_{m_{i}}\right)\frac{\textit{\textsf{C}}_{N-D_{m_{I+1}}}^{\alpha-1}}{M\textit{\textsf{C}}_{N-1}^{\alpha-1}}\\
&+\min_{m\in\mathcal{M}}\sum\limits_{n=2}^{\min\{D_{m},\alpha\}}
\frac{\textit{\textsf{C}}_{D_{m}}^{n}\textit{\textsf{C}}_{N-D_{m}}^{\alpha-n}}{M\textit{\textsf{C}}_{N-1}^{\alpha-1}}+\min\limits_{m\in\mathcal{M}}
      \frac{\textit{\textsf{C}}_{N-D_{m}}^{\alpha}}{M\textit{\textsf{C}}_{N-1}^{\alpha-1}},
\end{split}
\end{equation}
and $D_{m_{1}}\geq D_{m_{2}}\geq\cdots\geq D_{m_{I+1}}\geq D_{m}$ for any $m\in\mathcal{M}\setminus\{m_{1},...,m_{I+1}\}$ with $I$ obtained by $N\geq \sum\limits_{i=1}^{I}D_{m_{i}}$ and $N<\sum\limits_{i=1}^{I+1}D_{m_{i}}$.
\end{corollary}
\subsection{Special Case of $R^{*}$ for Uniform Requests}\label{sec:DRAnalyses:subsec:WorstDeliveryRateforSpecialCases}
\par We analyze $R^{*}$ for the special case that each user-group has the same number of distinct requests by assuming $D_{1}=\cdots=D_{M}=L$ with $1\leq L\leq N$. Based on \emph{Corollary} \ref{cor:WorstDeliveryRate_Case2}, we have the following on the worst delivery $R^{*}$ for $2\leq L\leq N$~and $ML>N$,
\begin{equation}\label{equ:cor:WorstDeliveryRate_Case2_UniformRequests}
\begin{split}
 R^{*}=&N-\left(G+\left(\frac{N}{M\alpha}-G\right)^{+}\right),
      \end{split}
\end{equation}
where
\begin{equation}\label{equ:ParametersoftheEquation_Uniformrequests}
   G=\frac{\left(N-L\right)\textit{\textsf{C}}_{N-L}^{\alpha-1}}{M\textit{\textsf{C}}_{N-1}^{\alpha-1}}+\frac{N}{M\alpha}.
\end{equation}
\par According to $G$ given in (\ref{equ:ParametersoftheEquation_Uniformrequests}), $R^{*}$ can be simplified as $R^{*}=N-G$. Then combining with \emph{Corollary} \ref{cor:WorstDeliveryRate_Case1} we have the following on $R^{*}$ for the uniform-request case.
\begin{corollary}\label{cor:WorstDeliveryRate_SpecialCase_URC}
Consider an $(N,M,D,\alpha)$ caching network with $D=\sum\limits_{m=1}^{M} D_{m}$ and $C=\frac{N}{M\alpha}$, where $D_{m}$ denotes the number of distinct requests of user-group $m\in \mathcal{M}$ and $C$ denotes the size of each cache. For $D_{1}=\cdots=D_{M}=L$, the worst delivery rate $R^{*}$ is given by
\begin{equation}\label{equ:cor:WorstDeliveryRate_SpecialCase_URC}
 R^{*}=\begin{cases}N-\frac{N(N+1)}{(\alpha+1)M},~L=1~\mathrm{and}~M\geq N;\\
  N-\frac{\left(N-L\right)\textit{\textsf{C}}_{N-L}^{\alpha-1}}{M\textit{\textsf{C}}_{N-1}^{\alpha-1}}-\frac{N}{M\alpha},~2\leq L\leq N~\mathrm{and}~L>\frac{N}{M}.\\
 \end{cases}
\end{equation}
\end{corollary}
\par Since we consider the worst delivery rate at the point that all the $N$ files are requested by the users and our coded prefetching algorithm can degenerate into the uncoded prefetching one by letting $\alpha=1$, a comparison of the delivery rate with those of traditional one-user-per-cache case \cite{Maddah-Ali2014,Ji2014,Sengupta2017} using uncoded prefetching at $D=ML\geq N$ can be performed. According to \cite[\emph{Theorem} 4]{Ji2014} and \cite[\emph{Lemma} 1]{Sengupta2017}, the achievable delivery rate for uncoded prefetching at $C=\frac{N}{M}$ and $D_{1}=\cdots=D_{M}=L$ is
\begin{equation}\label{equ:ConvnetionalDeliveryRate_SpecialCase_URC}
R_{\mathrm{URC}}^{\mathrm{Uncod}}=\begin{cases}\frac{M-1}{2},~L=1~\mathrm{and}~N\leq M\leq2N;\\
\min\left\{\frac{L(M-1)}{2},N-\frac{N}{M}\right\},~\mathrm{otherwise},\end{cases}
\end{equation}
whereas according to \emph{Corollary} \ref{cor:WorstDeliveryRate_SpecialCase_URC}, the worst delivery rate $R^{*}$ for uncoded prefetching at $C=\frac{N}{M}$ and $D_{1}=\cdots=D_{M}=L$ is
\begin{equation}\label{equ:cor:WorstDeliveryRate_SpecialCase_URC_alpha1}
 R^{*}=\begin{cases}N-\frac{N(N+1)}{2M},~L=1~\mathrm{and}~M\geq N;\\
  N-\frac{2N-L}{M},~2\leq L\leq N~\mathrm{and}~L>\frac{N}{M}.\\
 \end{cases}
\end{equation}
\par Comparing (\ref{equ:ConvnetionalDeliveryRate_SpecialCase_URC}) with (\ref{equ:cor:WorstDeliveryRate_SpecialCase_URC_alpha1}), we can easily prove that $R^{*}\leq R_{\mathrm{URC}}^{\mathrm{Uncod}}$ at $L=1$ and at $L\geq2$ for $D=LM\geq2N-L$. Such condition is obtained by $R_{\mathrm{URC}}^{\mathrm{Uncod}}-R^{*}=\frac{L(M-1)}{2}-N+\frac{2N-L}{M}=\frac{(M-2)(LM+L-2N)}{2M}\geq0$. Thus our delivery algorithm can outperform conventional uncoded prefetching for the uniform-request case at cache size $C=\frac{N}{M}$.
\begin{figure}[ht!]
   \subfigure[]
  {\begin{minipage}{1\textwidth}
  \centering
  \includegraphics[scale=0.62]{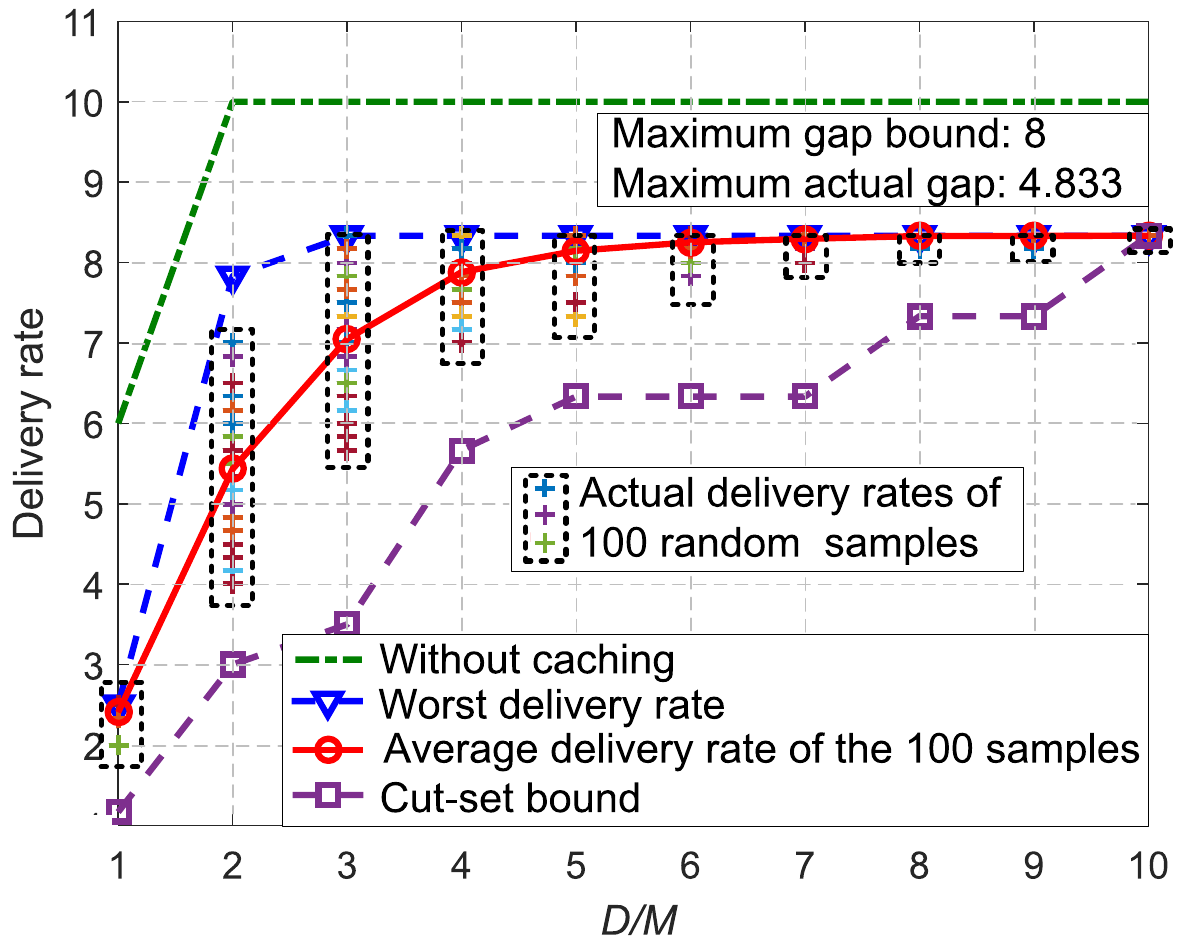}\label{fig:subfig:Fig2_DRSim_a}
  \end{minipage}}\\
    \subfigure[]
   {\begin{minipage}{1\textwidth}
   \centering
   \includegraphics[scale=0.62]{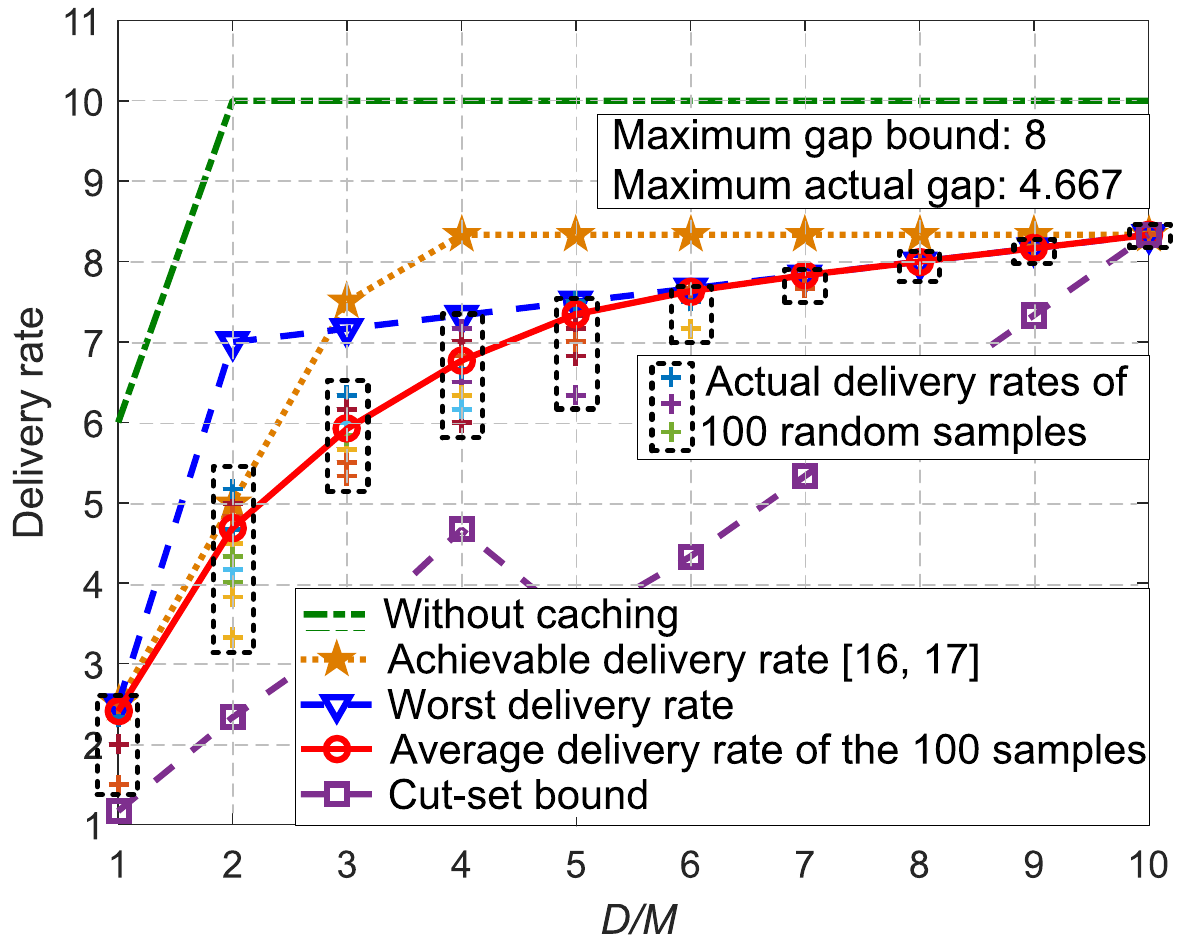}\label{fig:subfig:Fig2_DRSim_b}
   \end{minipage}}\\
       \subfigure[]
   {\begin{minipage}{1\textwidth}
   \centering
   \includegraphics[scale=0.62]{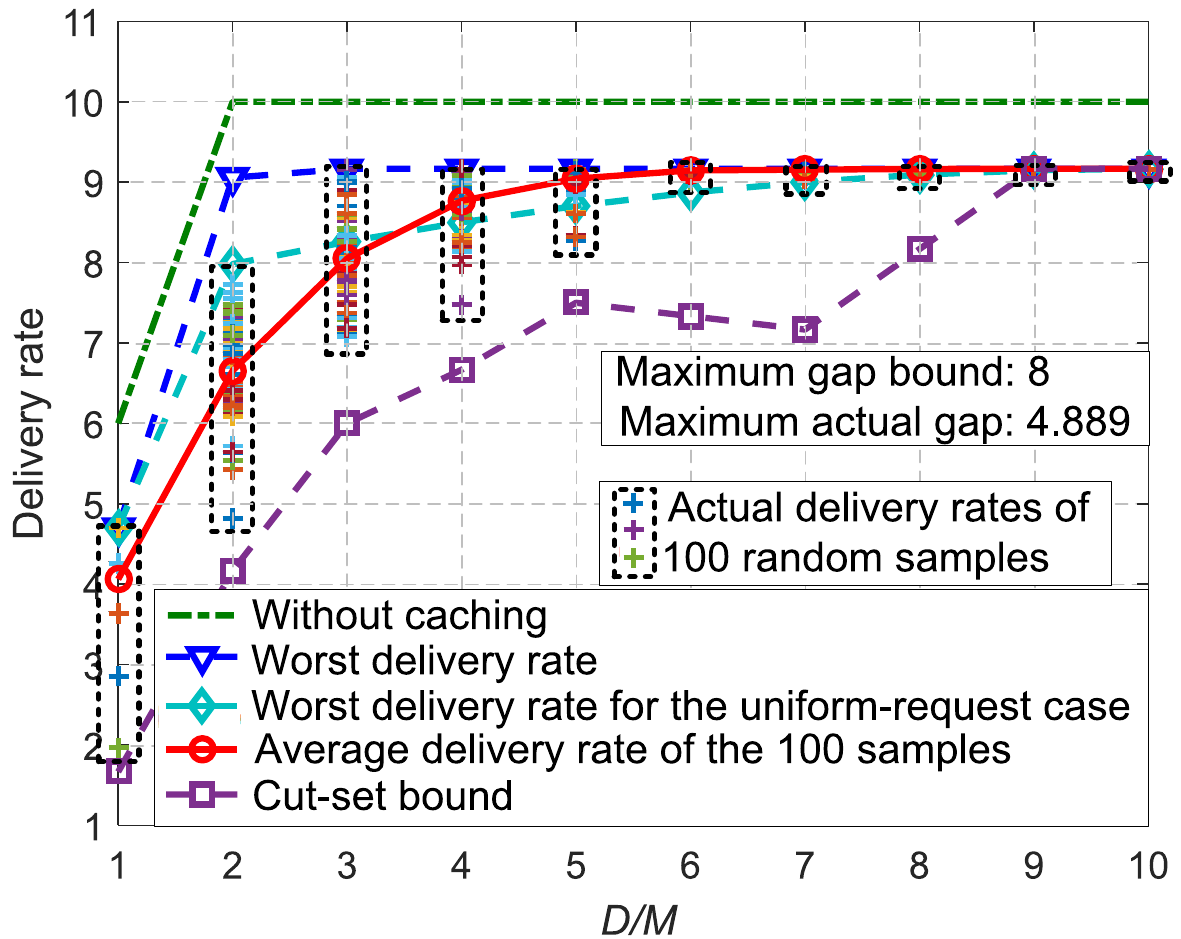}\label{fig:subfig:Fig2_DRSim_c}
   \end{minipage}}
\caption{Results of the delivery rate v.s. $\frac{D}{M}$, where $N=10$ and $M=6$. (a) Uncoded prefetching ($\alpha=1$) for arbitrary requests. (b) Uncoded prefetching ($\alpha=1$) for uniform requests such that $L=\frac{D}{M}$. (c) Coded prefetching ($\alpha=2$) for arbitrary requests.}\label{fig:Fig2_DeliveryRateAnalyses}
  \end{figure}
\section{Numerical Results}\label{sec:NumResults}
\par In this section, we present some numerical results for the achieved delivery rates given in Section \ref{sec:Analyses}. As a benchmark, the cut-set lower bound on the delivery rate of our model is computed, which, based on the derivation of the bound for traditional one-request-per-cache networks given in \cite[\emph{Theorem} 2]{Maddah-Ali2014}, can be formulated as
\begin{equation}\label{equ:LowerBound_CSB}
\begin{split}
 R\geq R_{\mathrm{CSB}}(N_{\mathrm{R}})&=\max\limits_{s,\{D_{m_{i}}\}_{i=1}^{s}} \sum_{i=1}^{s}D_{m_{i}}-\frac{sC}{\left\lfloor\vphantom{\sum\limits_{i=1}^{s}D_{m_{i}}}\right.\frac{N_{\mathrm{R}}}{\sum\limits_{i=1}^{s}D_{m_{i}}}\left\rfloor\vphantom{\sum\limits_{i=1}^{s}D_{m_{i}}}\right.},
\end{split}
\end{equation}
where $s\in\Big\{1,2,...,\min\{\lceil\frac{N_{\mathrm{R}}}{\min\limits_{m\in\mathcal{M}} D_{m}}\rceil,M\}\Big\}$ and $\sum\limits_{i=1}^{s}D_{m_{i}}\leq N_{\mathrm{R}}$. Setting $s=1$ we have $R_{\mathrm{CSB}}(N_{\mathrm{R}})\geq\max\limits_{m\in\mathcal{M}}\left(D_{m}-C/\lfloor\frac{N_{\mathrm{R}}}{D_{m}}\rfloor\right)$ based on (\ref{equ:LowerBound_CSB}). According to (\ref{equ:cor:WorstDeliveryRate_Case2}), we can upper bound the worst delivery rate $R^{*}(N)$ for the multi-request case $D>M$ as $R^{*}(N)=N-\left(G+\left(C-G\right)^{+}\right)\leq N-C$. Thus the gap between $R^{*}(N)$ and $R_{\mathrm{CSB}}(N_{\mathrm{R}})$ for any $N_{\mathrm{R}}\leq N$ can be upper bounded by
\begin{equation}\label{equ:Gapbound}
\begin{split}
  R^{*}(N)-R_{\mathrm{CSB}}(N_{\mathrm{R}})&\leq N-C-\max\limits_{m\in\mathcal{M}}\left(D_{m}-C\left/\vphantom{\left\lfloor\frac{N_{\mathrm{R}}}{D_{m}}\right\rfloor}\right.\left\lfloor\frac{N_{\mathrm{R}}}{D_{m}}\right\rfloor\right)\\
  &\leq N-C-\max\limits_{m\in\mathcal{M}}\left(D_{m}-C\right)\\
  &=N-\max\limits_{m\in\mathcal{M}}D_{m}.
\end{split}
\end{equation}
In simulations, the actual delivery rates for 100 groups of randomly produced requests for each given $D$ are calculated, and the corresponding average delivery rates are plotted. The worst delivery rate is determined by the maximum value of the worst delivery rates among the 100 groups, whereas the cut-set bound is determined by the minimum value of the cut-set bounds among the 100 groups.
 \begin{figure}[ht!]
   \subfigure[]
  {\begin{minipage}{0.495\textwidth}
  \centering
  \includegraphics[scale=0.62]{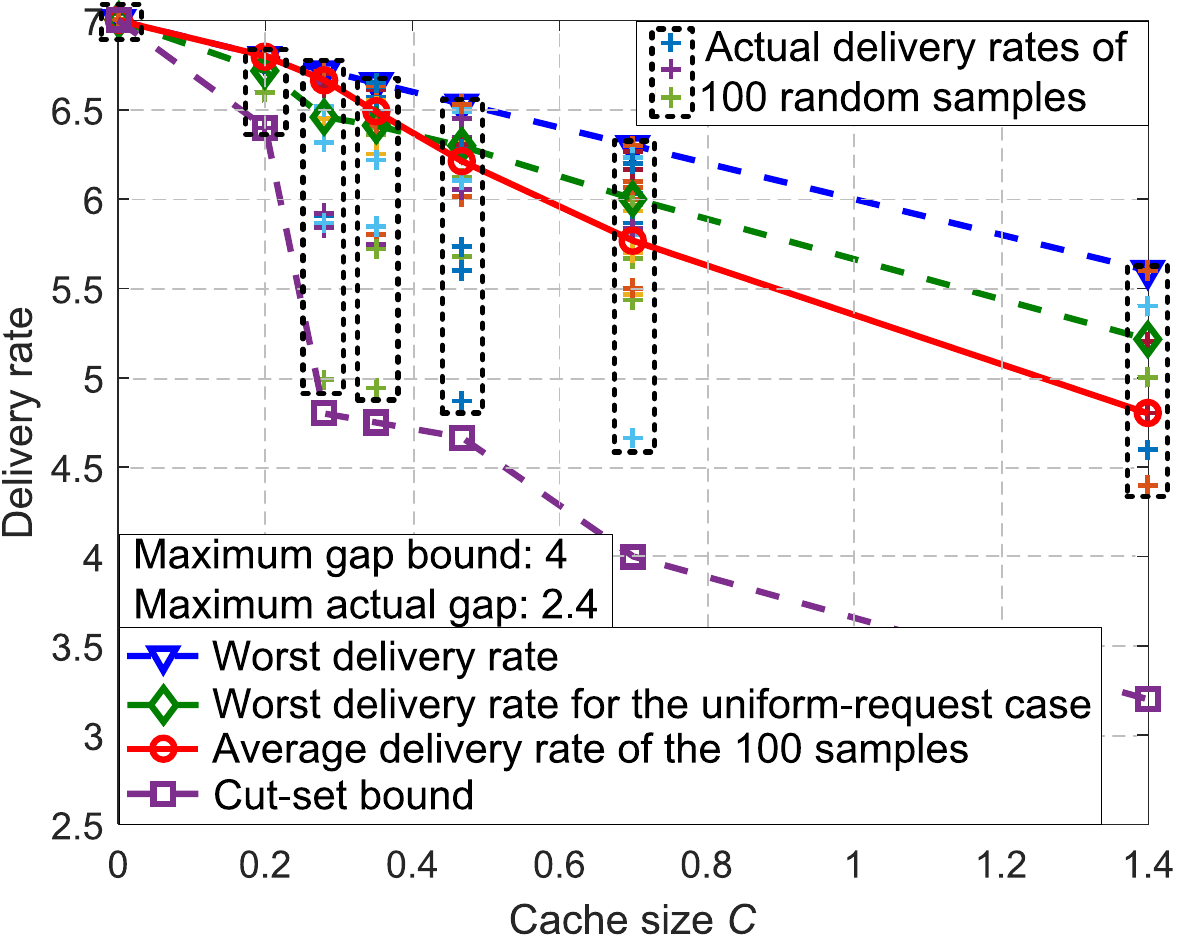}\label{fig:subfig:Fig3_RMSim_a}
  \end{minipage}}
      \subfigure[]
  {\begin{minipage}{0.495\textwidth}
  \centering
  \includegraphics[scale=0.62]{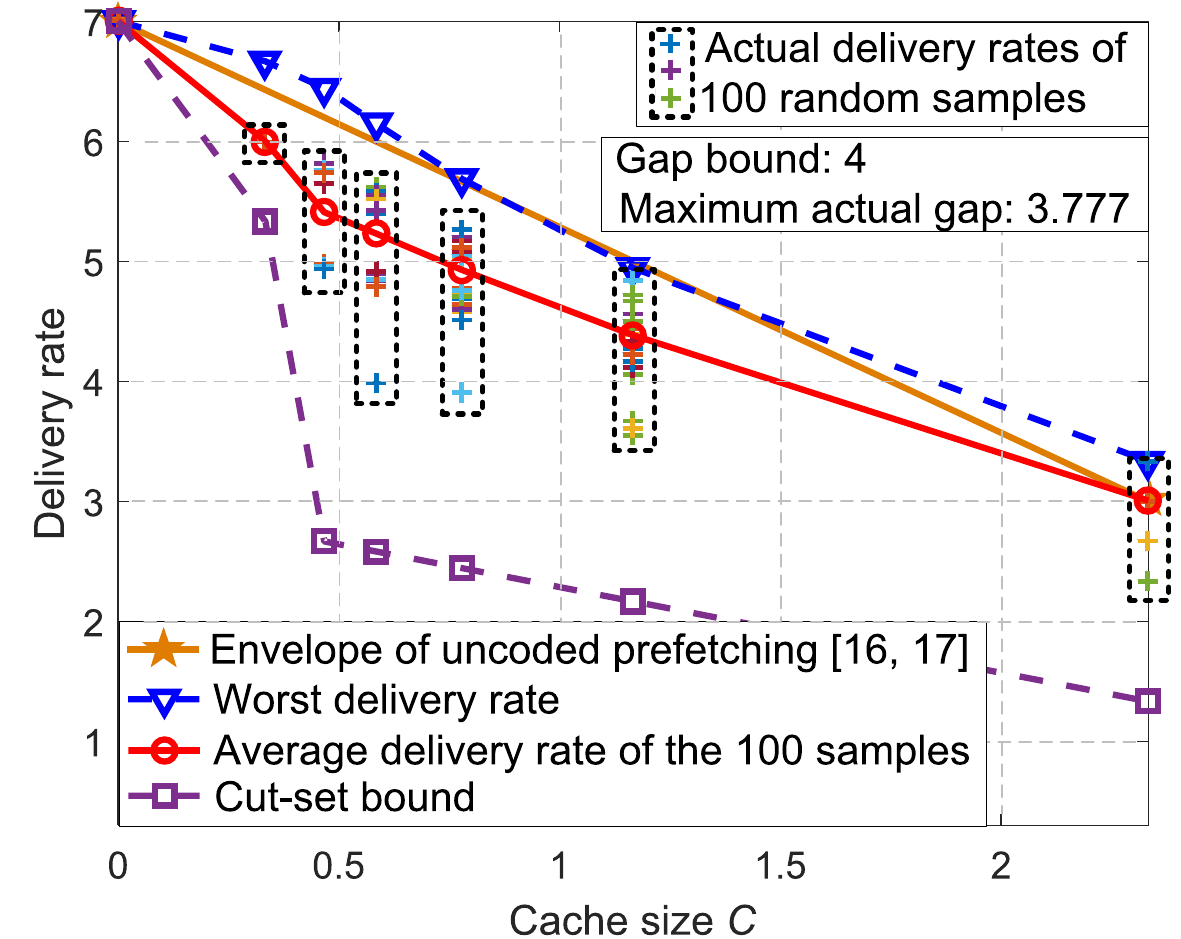}\label{fig:subfig:Fig3_RMSim_b}
  \end{minipage}}
    \subfigure[]
   {\begin{minipage}{0.495\textwidth}
   \centering
   \includegraphics[scale=0.62]{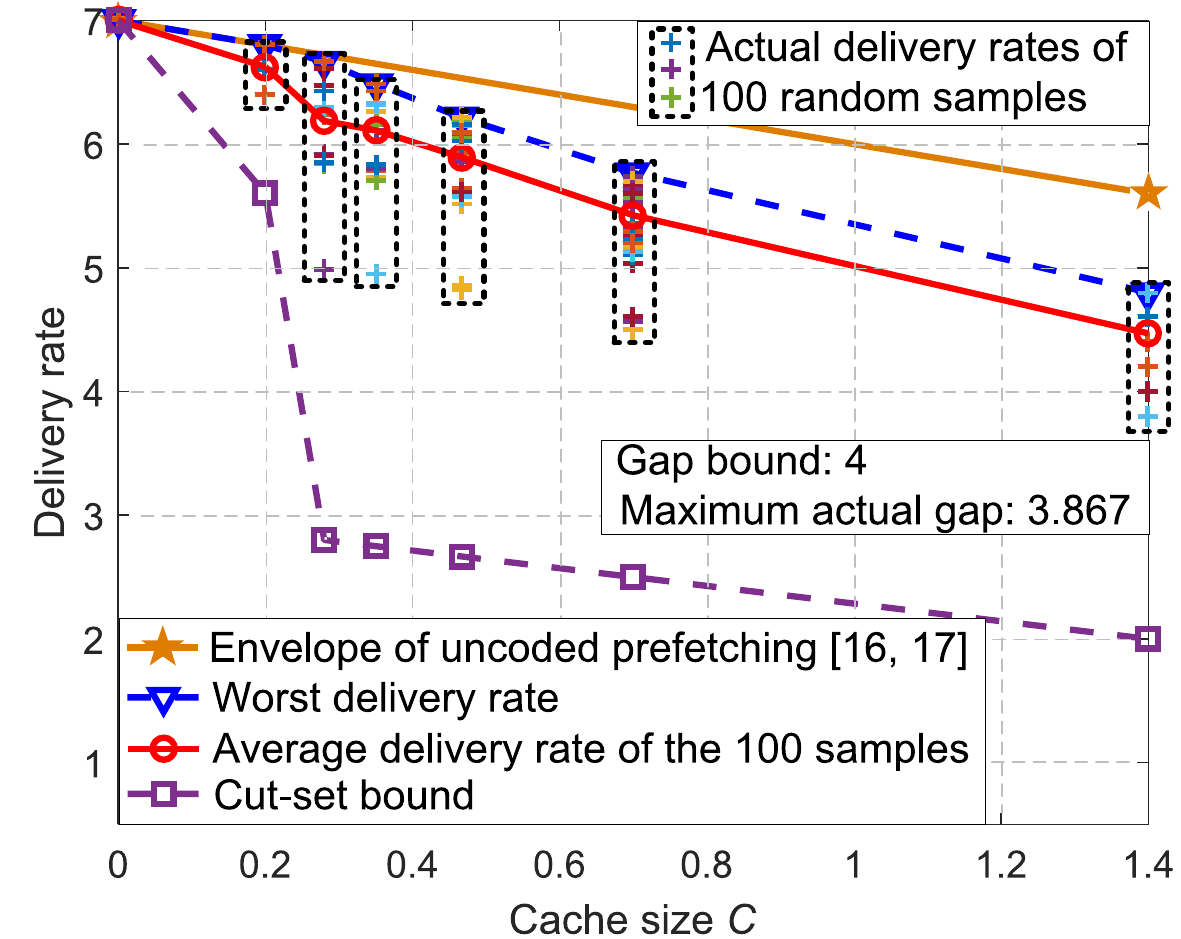}\label{fig:subfig:Fig3_RMSim_c}
   \end{minipage}}
       \subfigure[]
   {\begin{minipage}{0.495\textwidth}
   \centering
   \includegraphics[scale=0.62]{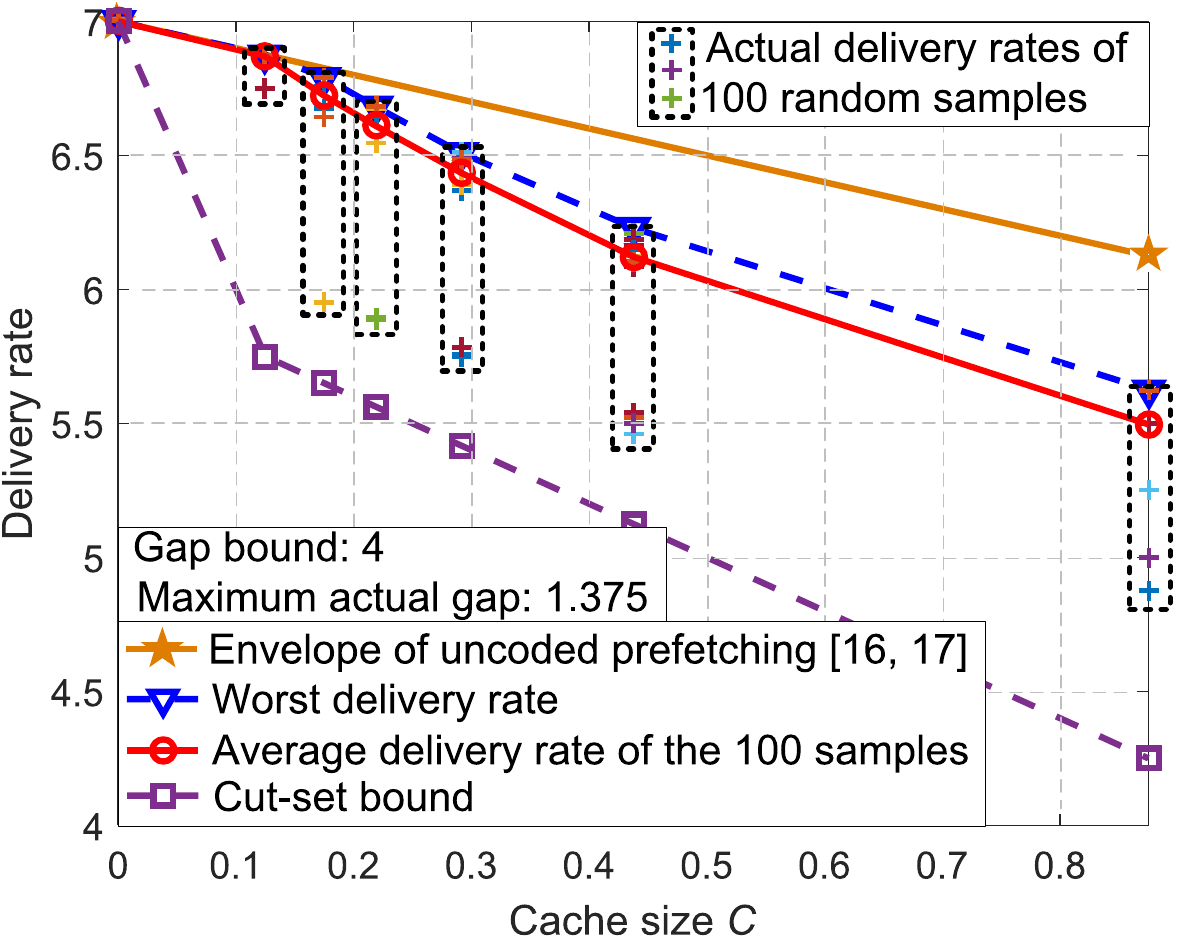}\label{fig:subfig:Fig3_RMSim_d}
   \end{minipage}}
 \caption{Results of rate-memory pairs over $0\leq C\leq \max\limits_{\alpha}\frac{N}{M\alpha}=\frac{N}{M}$, where $N=7$. (a) Arbitrary multiple requests with $M=5$ and $D=15$. Uniform multiple requests with $L=3$ and (b) $M=3<N$. (c) $M=5<N$. (d) $M=8>N$.}\label{fig:Fig3_RateMemoryAnalyses}
  \end{figure}
\par For the proposed delivery scheme, Fig. \ref{fig:Fig2_DeliveryRateAnalyses} shows detailed results of the delivery rate versus the ratio of the total request number $D$ to the cache number $M$, i.e., $\frac{D}{M}$, for both uncoded and coded prefetching. Fig. \ref{fig:Fig3_RateMemoryAnalyses} shows detailed results of rate-memory pairs for both arbitrary and uniform multiple requests. It can be seen that all the actual and average delivery rates in the seven subfigures fall between the proposed worst delivery rate and the lower bound, with the gap lower than the bound given in (\ref{equ:Gapbound}). Furthermore, it can be seen that both the average and worst delivery rates increase with the sum number of requests $D$ according to Fig. \ref{fig:Fig2_DeliveryRateAnalyses}, and decrease with the cache size $C$ according to Fig. \ref{fig:Fig3_RateMemoryAnalyses}. Note that the cut-set bound may not monotonously change as shown in Figs. \ref{fig:subfig:Fig2_DRSim_b} and \ref{fig:subfig:Fig2_DRSim_c} as it is chosen from 100 independent random samples for each point under the minimax criterion.
\par Additionally, Fig. \ref{fig:subfig:Fig2_DRSim_b} indicates that our proposed delivery scheme for the uniform-request case can achieve the same or a lower average delivery rate than that given in \cite{Ji2014,Sengupta2017}. It can achieve the same or a lower worst delivery rate than that given in \cite{Ji2014,Sengupta2017} except the point $\frac{D}{M}=2$, which is consistent with the analysis provided in Section \ref{sec:DRAnalyses:subsec:WorstDeliveryRateforSpecialCases} since $D=M$ or $D=ML\geq2N-L$ for $D>M$ holds at all other points. Under such conditions, whether $M\geq N$ or not, our proposed coded prefetching can achieve a better rate-memory trade-off than the uncoded prefetching \cite{Ji2014,Sengupta2017} over $0\leq C\leq\frac{N}{M}$, as shown in Figs. \ref{fig:subfig:Fig3_RMSim_c} and \ref{fig:subfig:Fig3_RMSim_d}. Although the worst delivery rates plotted in Fig. \ref{fig:subfig:Fig3_RMSim_b} are not all lower than the envelope of the delivery rate provided by \cite{Ji2014,Sengupta2017} as $LM=9<2N-L=11$, the average delivery rats are lower.
\begin{figure}[ht!]
   \subfigure[]
   {\begin{minipage}{0.495\textwidth}
   \centering
   \includegraphics[scale=0.62]{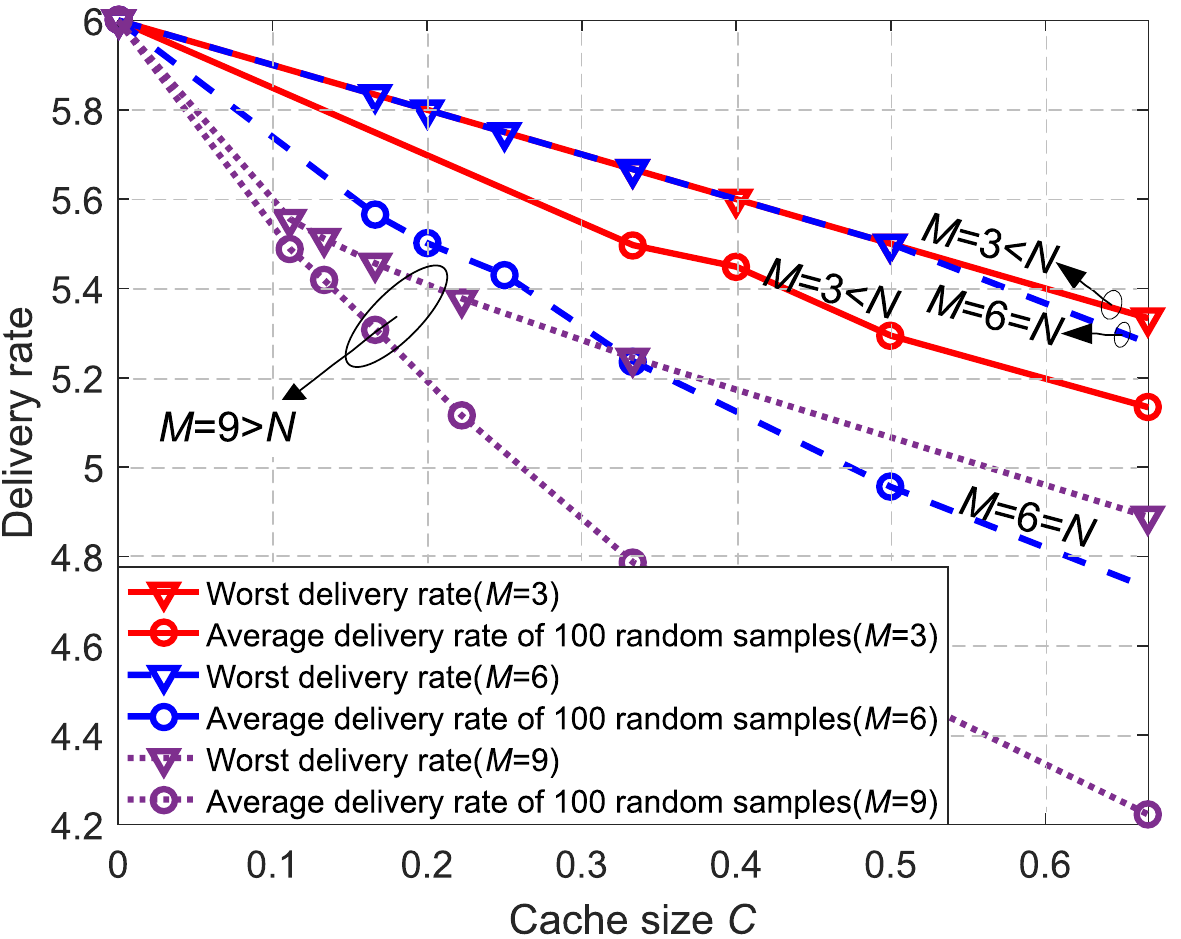}\label{fig:subfig:Fig4_RMCompM_a}
   \end{minipage}}
       \subfigure[]
   {\begin{minipage}{0.495\textwidth}
   \centering
   \includegraphics[scale=0.62]{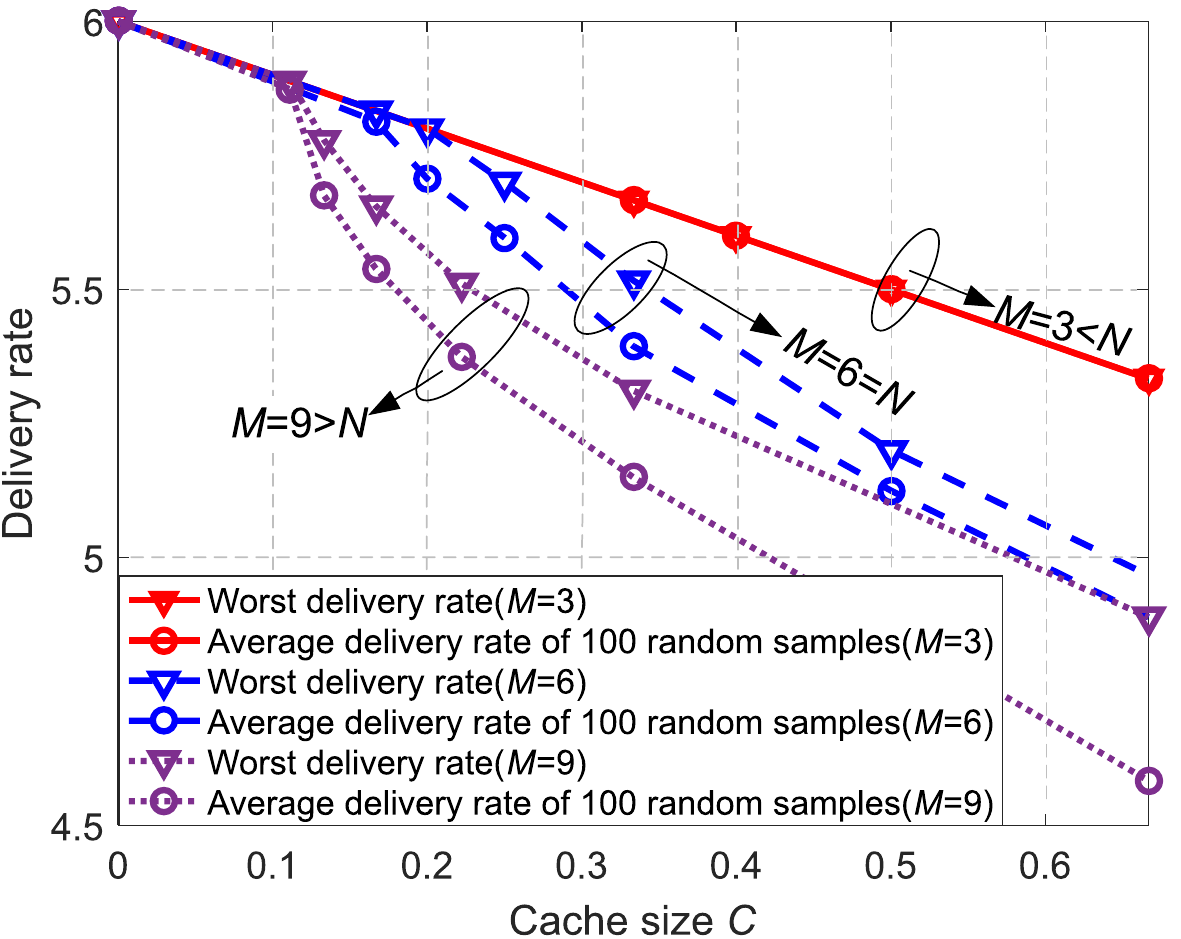}\label{fig:subfig:Fig4_RMCompM_b}
   \end{minipage}}
\caption{Results of rate-memory pairs for different $M$ over $0\leq C\leq \min\limits_{M}\frac{N}{M}$, where $N=6$. (a) Arbitrary multiple requests with $D=10$. (b) Uniform multiple requests with $D=18$.}\label{fig:Fig4_RateMemoryComparisonM}
  \end{figure}
\par Finally, Fig. \ref{fig:Fig4_RateMemoryComparisonM} shows the results of rate-memory pairs under different $M$ for both arbitrary and uniform multiple requests, which indicates that increasing $M$ can achieve better rate-memory trade-offs since both the average and worst delivery rates decrease with $M$.
\section{Conclusions}\label{sec:Conclusion}
\par We have considered a centralized caching network, where a server serves several groups of users, each having a common shared homogeneous fixed-size cache and requesting arbitrary multiple files. An efficient file delivery scheme with explicit constructions by the server to meet the multi-requests of all user-groups was first proposed. Then the rate as well as the worst rate of the proposed delivery scheme were analyzed. We showed that our caching model and delivery scheme can incorporate some existing coded caching schemes as special cases. Moreover, for the special case of uniform requests and uncoded prefetching, we made a comparison with existing results, and showed that our approach achieved a lower delivery rate. Finally, numerical results demonstrated the effectiveness of the proposed delivery scheme.

\appendices
\section{Search Algorithm for the $(\alpha+1)$-request Sets}\label{app:SearchingAlgorithm}
\par The search algorithm of the valid $(\alpha+1)$-request sets $\mathcal{V}$ consists of two steps:
\par \emph{Step 1:} Identify every $\alpha+1$ different user-groups to find $\mathcal{V}$ satisfying (\ref{equ:PacketofType4_ConsiditionforStep1_a}) based on a reference user-group $m_{0}$ for $m_{0}=1,...,M-\alpha$. Each time given the reference user-group $m_{0}$ and one of its requests $n_{m_{0}}$, the server arbitrarily picks out $\alpha$ distinct requests $n_{m_{1}}$, $n_{m_{2}},...,n_{m_{\alpha}}$ from other $\alpha$ user-groups $m_{1},...,m_{\alpha}$ from $\{m_{0}+1,...,M\}$ to guarantee that $\mathcal{V}=\{n_{m_{0}}, n_{m_{1}},...,n_{m_{\alpha}}\}$ satisfies (\ref{equ:PacketofType4_ConsiditionforStep1_a}).
\par \emph{Step 2:} Make sure that each $\mathcal{V}$ corresponds to a totally different packet-group based on (\ref{equ:PacketofType4_ConsiditionforStep1_b}). As given a reference user-group $m_{0}$, there will be more than one choice of $n_{m_{0}}$, or given $n_{m_{0}}$ or different $n_{m_{0}}$ from different reference user-groups, there will be more than one choice of the request sets, resulting in at least two obtained ($\alpha+1$)-request sets with the same $\alpha$ requests. This may lead to different packet-groups containing a common Type-\MyRoman{4} packet. Thus (\ref{equ:PacketofType4_ConsiditionforStep1_b}) is added with (\ref{equ:PacketofType4_ConsiditionforStep1_a}) for any two obtained ($\alpha+1$)-request sets to guarantee different request sets for totally different packet-groups.
\par Based on (\ref{equ:PacketofType4_ConsiditionforStep1_a}) and (\ref{equ:PacketofType4_ConsiditionforStep1_b}), all the valid ($\alpha+1$)-request sets can be obtained. The corresponding search procedure is summarized in \emph{Algorithm} \ref{alg:SerachingRequestSets} and the illustration of the search algorithm is presented in \emph{Example} \ref{exm:IllustrationofSearchignAlgorithm}.
\begin{example}\label{exm:IllustrationofSearchignAlgorithm}
 An example is shown in Table \ref{tab:DeliverySchemeofPacketType4_Step1}, from which 7 valid ($\alpha+1$)-request sets satisfying both (\ref{equ:PacketofType4_ConsiditionforStep1_a}) and (\ref{equ:PacketofType4_ConsiditionforStep1_b}) can be obtained. The sets satisfying (\ref{equ:PacketofType4_ConsiditionforStep1_a}) but not selected are $\{1,4,5,7\}$, $\{1,4,7,8\}$, $\{1,5,7,8\}$, $\{2,5,6,8\}$, $\{3,4,6,8\}$ and $\{4,5,7,8\}$. For example, set $\{2,5,6,8\}$ is not selected since $\{1,5,6,8\}$ and $\{2,5,6,8\}$ do not meet (\ref{equ:PacketofType4_ConsiditionforStep1_b}) and thus only one set is selected.
\end{example}
  \begin{table*} [ht]
  \setcaptionwidth{5.3in}
  \centering
\begin{tabular}{|c|c|c|c|c|c|c|c|  }
\hline
  \multicolumn{2}{|c|}{}&\multicolumn{6}{c|}{$\mathcal{M}$} \\
 \hline
   \multicolumn{2}{|c|}{User-group $m$}& 1 &2 &3 &4&5&6  \\
   \hline
     \multicolumn{2}{|c|}{Requests}& $\{1,2,3\}$ & $(2,4)$ & $\{2,3\}$
     & $\{3,5\}$ &$\{6,7\}$&$\{8\}$\\
   \hline
   \multirow{8}{*}{($\alpha+1$)-request sets}&\multirow{6}{*}{$m_{0}=1$}& \multicolumn{6}{c|}{ $\mathcal{V}=\{1,4,5,6\}$ when $m_{1}=2$, $m_{2}=4$,
  $m_{3}=5$}\\
  && \multicolumn{6}{c|}{ $\mathcal{V}=\{1,4,5,8\}$ when $m_{1}=2$, $m_{2}=4$, $m_{3}=6$}\\
   && \multicolumn{6}{c|}{$\mathcal{V}=\{1,4,6,8\}$ when $m_{1}=2$,
  $m_{2}=5$, $m_{3}=6$}\\
   && \multicolumn{6}{c|}{$\mathcal{V}=\{1,5,6,8\}$ when $m_{1}=4$, $m_{2}=5$, $m_{3}=6$}\\
      && \multicolumn{6}{c|}{$\mathcal{V}=\{2,5,7,8\}$ when $m_{1}=4$, $m_{2}=5$, $m_{3}=6$}\\
      && \multicolumn{6}{c|}{$\mathcal{V}=\{3,4,7,8\}$ when $m_{1}=2$, $m_{2}=5$, $m_{3}=6$}\\
     \cline{2-8}
     &$m_{0}=2$&\multicolumn{6}{c|}{ $\mathcal{V}=\{4,5,6,8\}$ when $m_{1}=4$, $m_{2}=5$, $m_{3}=6$}\\
     \cline{2-8}
     &$m_{0}=3$&\multicolumn{6}{c|}{ None}\\
     \hline
\end{tabular}\caption{Illustration of searching the ($\alpha+1$)-request sets, where $\alpha=3$.}\label{tab:DeliverySchemeofPacketType4_Step1}
\end{table*}

\ifCLASSOPTIONcaptionsoff
  \newpage
\fi

\bibliographystyle{IEEEtran}
\end{document}